%% file: MainDocument.tex
\documentclass[prodmode]{article} 

\usepackage[ruled]{algorithm2e}
\usepackage{setspace}
\usepackage{fullpage}
\usepackage{authblk}
\def\subcaptionfont{\fontsize{8}{10}\selectfont}%
\def\subcaption#1{{\centering\subcaptionfont#1\par}}

\usepackage{comment}

\usepackage{listings}

\usepackage{amsmath, amsfonts, amsthm}
\newcommand{\eat}[1]{}

\input{sections/defs.tex}

\SetAlFnt{\small}
\SetAlCapFnt{\small}
\SetAlCapNameFnt{\small}
\SetAlCapHSkip{0pt}
\IncMargin{-\parindent}

\begin{document}

\title{Efficient Atlasing and Search of Configuration Spaces of Point-Sets Constrained by Distance Intervals.}

\eat{
\author[1]{Aysegul Ozkan, Rahul Prabhu, Troy Baker, James Pence, Jorg Peters and Meera Sitharam}
\affil[1]{Department of Computer \& Information Science \& Engineering, University of Florida }
}

\author{Aysegul Ozkan, Rahul Prabhu, Troy Baker, James Pence, Jorg Peters and Meera Sitharam
\thanks{Department of Computer \& Information Science \& Engineering, University of Florida }
}

\eat{
\author[1]{Aysegul Ozkan}
\author[1]{Rahul Prabhu} 
\author[1]{Troy Baker} 
\author[1]{James Pence}
\author[1]{Jorg Peters}
\author[1]{Meera Sitharam} 
\affil[1]{Department of Computer \& Information Science \& Engineering, University of Florida }
}

\maketitle
\input{sections/Abstract}

\input{sections/Introduction}
\input{sections/Theory}
\input{sections/AlgorithmicIdeas}

\input{sections/Results}

\input{sections/Architecture}
\input{sections/Conclusion}

\bibliographystyle{plain}
\bibliography{easal,stickysphere,nigms,jorg,Dmay04,intro}



\medskip

\end{document}

%% file: sections/defs.tex
\newtheorem{theorem}{Theorem}

\usepackage{url}
\usepackage{enumerate}
\usepackage{float}
\usepackage{flafter}
\usepackage{psfrag}
\usepackage{overpic}
\usepackage{subfigure}
\usepackage{caption}
\usepackage{multirow}
\usepackage{array}

\usepackage{amsmath}

\usepackage{epsfig}

\usepackage{color}

\newcommand{\fig}{./sections/fig}
\newcommand{\comm}[1]{}

\def\IL{{\it left}}

\def\IR{{\it right}}

\newcommand{\figref}[1]{Fig.~\ref{#1}}
\newcommand{\exref}[1]{Ex.~#1}

\newcolumntype{C}[1]{>{\centering\let\newline\\\arraybackslash\hspace{0pt}}m{#1}}

\newcommand{\EASAL}{\texttt{EASAL}}

\newcommand{\distall}{$C_1$}
\newcommand{\distexist}{$C_2$}

\newcommand{\toytwod}{Toy3}

\newcommand{\toyhelix}{6Atom}
\newcommand{\bighelix}{Toy20}
\newcommand{\ctwo}{\ref{eqn:preferredConstraints}}
\newcommand{\cone}{\ref{eqn:constraints}}

\definecolor{brown}{RGB}{150,50,50}
\definecolor{pink}{RGB}{219, 48, 122}

\definecolor{seagreen}{RGB}{48,178,139}
\definecolor{yelloworange}{RGB}{200,146,0}
\definecolor{purple}{RGB}{123,104,238}
\definecolor{grayoned}{RGB}{0,0,255}
\definecolor{green2d}{RGB}{0,150,0}
\definecolor{plum}{RGB}{127,0,126}

%% file: sections/Abstract.tex
\begin{abstract}

For configurations of point-sets that are pairwise constrained by distance
intervals, the EASAL software implements a suite of algorithms that
characterize the structure and geometric properties of the configuration
	space\let\thefootnote\relax\footnotetext{ \noindent This research was
	supported in part by NSF DMS-0714912, a University of Florida
	computational biology seed grant, NSF CCF-1117695, NSF DMS-1563234, NSF
	DMS-1564480

\textbf{Note:}
\begin{enumerate}
\item The EASAL software has two versions. The TOMS submission contains only
the backend of EASAL, without GUI and with text input and output. The
experimental results in Section \ref{sec:atlasandpath} of the paper, can be
reproduced with this version using the sample input files given in the files
directory. See Section 5 of the included ‘TOMSUserGuide.pdf’ for detailed
instructions on how to run the test driver. An optional GUI (not part of TOMS
submission) which can be used for intuitive visual verification of the results,
can be found at the EASAL repository.  Instructions on how to install and how
to use and major functionalities offered by the GUI are detailed in the
‘CompleteUserGuide’ found in the bitbucket respository which can be found at
\url{http://bitbucket.org/geoplexity/easal} \cite{easalSoftware}.

\item A video presenting the theory, applications, and software components of
EASAL is available at \url{http://www.cise.ufl.edu/~sitharam/EASALvideo.mpg}
\cite{easalVideo}.
\item A web version of the software can be found at
\url{http://ufo-host.cise.ufl.edu} (runs on Windows, Linux, and Chromebooks with the latest, WebGL 2.0 enabled
 google chrome or mozilla firefox web browsers).
\item EASAL screen shots and movies have been used in the papers
\cite{sym8010005,Ozkan2011,Wu2012,Ozkan2014MainEasal,Ozkan2014MC,Ozkan2014Jacobian,Wu2014Virus}
to illustrate definitions and theoretical results.
\end{enumerate}}.
The algorithms generate, describe and explore these configuration spaces using
generic rigidity properties, classical results for stratification of
semi-algebraic sets, and new results for efficient sampling by convex
parametrization. The paper reviews the key theoretical underpinnings, major
algorithms and their implementation. The paper outlines the main applications
such as the computation of free energy and kinetics of assembly of supramolecular
structures or of clusters in colloidal and soft materials. In addition, the
paper surveys select experimental results and comparisons.

\end{abstract}

%% file: sections/Introduction.tex
\section{Introduction}
\label{sec:intro}

We present a software implementation of the algorithm EASAL (Efficient Atlasing
and Search of Assembly Landscapes) \cite{Ozkan2011}. This implementation
generates, describes, and explores the feasible relative positions of two
point-sets $A$ and $B$ of size $n$ in $\mathbb{R}^3$ that are mutually
constrained by distance intervals. Formally, a Euclidean orientation-preserving
isometry $T \in SE(3)$ is \emph{feasible }
if, for $\emph{dist}_{a,b}$ defined as the Euclidean norm
$||a-T(b)||$, the following hold:

\begin{align}
\forall (a\in A, b\in B),\qquad& \emph{dist}_{a,b} \qquad \ge \rho_{a,b}& \tag{\distall}\label{eqn:constraints}\\
\exists (a\in A, b\in B),\qquad& \emph{dist}_{a,b} - \rho_{a,b} \le \delta_{a,b}, & \rho_{a,b}, \delta_{a,b} \in \mathbb{R}_+.\tag{\distexist}\label{eqn:preferredConstraints}
\end{align}

\noindent Constraint \cone\ means that $T$ is infeasible when there exists a
pair $(a, T(b))$ that is too close. Constraint \ctwo\ implies that at least one
pair $(a,T(b))$ is within a \emph{preferred} distance interval.  Consider for
example, sets $A$ and $B$ of centers of non-intersecting spheres (see
\figref{fig:pctreeInput} and \figref{fig:3Input}). With $\rho_a$,
$\rho_b$ the sphere radii, the constant $\rho_{a,b}$ in \cone\ equals $\rho_a +
\rho_b$.  Note that the ambient dimension of Problem (\cone, \ctwo) is 6,
namely, the dimension of SE(3).  When $T$ is feasible, the \emph{Cartesian
configuration} $T(B)$ is called a \emph{realization} of the constraint system
(\cone, \ctwo).  When  $\delta_{a,b}\approx 0$ the effective dimension of the
realization space is 5.

The input to EASAL consists of up to four components. 
\begin{enumerate}
\item [--] $k=2$ point-sets $A$ and $B$ with $n$ points each. (The submitted
implementation is for two point-sets, but the theory and the algorithms
generalize to $k$ point-sets and ambient dimension $6(k-1)$)

\item[--] The pairwise distance interval parameters $\rho_{a,b}, \delta_{a,b} \in
\mathbb{R}_+$.

\item[--] Optional: global constraints imposed on the overall configuration. 

\item[--] Optional: a set of active constraints of interest. (Only constraint
regions including at least one of these active constraints is sampled and added
to the atlas.)
\end{enumerate}

The main output of EASAL is the dimensional, topological and geometric
structure of the \emph{realization space}, i.e., all $T(B)$ satisfying (\cone,
\ctwo). The realization space is represented as the \emph{sweep} of the
individual realizations (see \figref{fig:pctreeSweep} and
\figref{fig:RealizationSweepView}). The sweep representation shows $A$ together with all
feasible realizations $T(B)$ traced out.

To describe this space, EASAL employs three strategies.  First, EASAL
partitions the realization space into \emph{active constraint regions}, each
defined by the set of \emph{active constraints}, i.e., the pairs $(a,b)$
satisfying \ctwo.  These pairs are edges of the \emph{active constraint graph}
used to label the region.  Such a graph can be analyzed by generic
combinatorial rigidity theory \cite{CombinatorialRigidity}, in particular, the
co-dimension of an active constraint region (see Section
\ref{sec:stratification}) is typically the number of active constraint edges.
Since the active constraint regions satisfy polynomial equations and
inequalities, the realization space is semi-algebraic set (a union of sets
defined by polynomial inequalities).  This is the setting of a Thom-Whitney
stratification of semi-algebraic sets \cite{Kuo}.

Second, EASAL organizes and represents the active constraint regions in a
partial order (directed acyclic graph) so that the active constraint graph of a
region is a subgraph of the active constraint graph of its boundary regions.
This organization is called the \emph{atlas}. To construct the atlas, EASAL
recursively starts from the interior of an active constraint region and locates
boundary regions of strictly one dimension less. Such boundary regions
generically have exactly one additional active constraint and the active
constraint graph has one additional edge. Considering only boundary regions of
exactly one dimension less improves robustness over searching directly for
lowest-dimensional regions. We note that, when a new \emph{child} region of one
dimension less is found, all its higher dimensional \emph{ancestor} regions are
immediately discovered since they correspond to a subset of the active
constraints.  Therefore, even if a region is missed at some stage, it will be
discovered once any of its descendants are found, for example, through one of its siblings.

Third, to locate the boundary region satisfying an additional active
constraint, EASAL applies the theoretical framework developed in
\cite{SiGa:2010}. EASAL efficiently maps (many to one) a $d$-dimensional active
constraint region $R$ with active constraint graph $G$, to a convex region of
$\mathbb{R}^d$ called the \emph{Cayley configuration space} of $R$. Define 
a \emph{non-edge} of $G$ as a pair $(a,b)$ not connected by an edge in $G$.
The Cayley configuration space of $R$ is defined intuitively as the set of
\emph{realizable} lengths of $d$ chosen non-edges of $G$. The variables representing these 
non-edge lengths are called the \emph{Cayley parameters}. In what follows, we simply refer to the
non-edges as Cayley parameters.  Since the Cayley
configuration space is convex, it allows for efficient sampling and search. In
addition, it is efficient to compute the inverse map from each point in the
Cayley configuration space (a \emph{Cayley configuration}) to its finitely many
corresponding Cartesian realizations.  The Cayley configuration space of a
$d$-dimensional active constraint region $R$ is discretized and represented as a
$d$-dimensional grid.  The Cayley points adjacent to the lower dimensional boundary regions of $R$ are
highlighted in different colors (See \figref{fig:pctreeSpace}).


Efficiency, accuracy, and tradeoff guarantees have been formally established for
EASAL (see Section \ref{sec:complexity}). The total number of active constraint
regions in the atlas could be as large as $O(k^2 \cdot n^{12k})$.  The maximum
dimension of a region is $6(k-1)$. If $r$ regions of dimension $d$ have to be
sampled, EASAL requires time linear in $r$ and exponential in $d$.  EASAL can
explore assemblies up to a million regions for \emph{small assemblies} in a few
hours on a standard laptop (see Section \ref{sec:atlasandpath}). By small
assemblies we mean constraint problems with $n \le 5000$ and $k=2$; or $n\le 3$
and $k\le 18$.  Efficiency can improve significantly when the point-sets are
identical, by exploiting symmetries in the configuration space
\cite{sym8010005}.

Section \ref{sec:results} surveys numerical experimental results from
\cite{Ozkan2014MainEasal}, for (i) generating the atlas, (ii) using the atlas
to find paths between active constraint regions and (iii) using the atlas to
find the neighbor regions of an active constraint region. We also survey
experimental results from \cite{Ozkan2014MC}, comparing the performance of
EASAL with Metropolis Markov chain Monte Carlo (MC) and from \cite{Wu2014Virus}
for EASAL predicting the sensitivity of icosahedral T=1 viruses towards
assembly disruption.\\

\noindent\textbf{Organization:} 
After briefly reviewing applications of EASAL to molecular and materials
modeling and related work, the remainder of the paper is organized as follows.
Section \ref{sec:theory} discusses the theory underlying EASAL. Section
\ref{algorithms} discusses the algorithmic ideas and implementation. Section
\ref{sec:results} surveys experimental results, Section \ref{architecture}
sketches the software architecture.

\subsection{Application to Molecular and Materials Modeling} 

EASAL provides a new approach to the longstanding challenges in molecular and
soft-matter self-assembly under short range potential interactions.  EASAL can
be used to estimate free-energy, binding affinity and kinetics. For example, EASAL
can be applied to (a) supramolecular self-assembly or docking starting from rigid
molecular motifs e.g., helices, peptides, ligands etc. or (b) self-assembly of
clusters of multiple particles each consisting of 1-3 spheres - e.g., in
amphiphiles, colloids or liquid crystals. 

In the context of molecular assembly, rigid components of the molecules
correspond to the input point-sets $A$ and $B$, and atoms correspond to the
points $a\in A$ and $b\in B$. The active constraint regions correspond to
regions of constant potential energy derived from discretized
Lennard-Jones~\cite{Jones463} potential energy terms. It is intractable or at
least prodigiously expensive to atlas large molecular assemblies by any naive
global method. Assemblies are typically recursively decomposed into smaller
assemblies (defined above) and recombined. Generally, the input molecules have
a small set of interfaces (pairs of atoms, one from each molecule) where bond
formation is feasible. These are given as input by specifying a set of active
constraints of interest corresponding to the interfaces. EASAL atlases only
those $r$ active constraint regions where at least one of these constraints is
active (i.e., $C_2$ holds).

\subsubsection{Geometrization of Molecular Interactions in EASAL} 
\label{sec:geometrization}
In EASAL, the inter-atomic Lennard-Jones potential energy terms are
\emph{geometrized} into 3 main regions: (i) large distances at which no force
is exerted between the atoms, such atom pairs, called inactive constraints,
correspond to $(a,b)$ such that $dist_{a,b} > \rho_{a,b} + \delta_{a,b}$, (ii)
very close distances that are prohibited by inter-atomic repulsion or
inter-atomic collisions and violating \cone. (iii) the interval between these,
known as the Lennard-Jones \emph{well}, in which bonds are formed,
corresponding to the preferred distance or active constraints defined in \ctwo.

The pairwise Lennard-Jones terms are typically input only for selected pairs of
atoms, one from each rigid component. Hard-sphere \emph{steric} constraints,
apply to all other pairs and enforce (i) and (ii) with $\delta_{a,b} = 0$ in
\ctwo.  Having more active constraints corresponds to lower potential energy,
as well as to lower effective dimension of the region. The lowest potential
energy is attained at zero-dimensional regions, i.e., for rigid active constraint
graphs and finitely many configurations. For each rigid active constraint graph
$G$, the corresponding potential energy \emph{basin} includes well-defined
portions of higher dimensional regions whose active constraint graphs are
non-trivial subgraphs of $G$. In this manner the Cartesian configuration space
is partitioned into potential energy basins. Free energy of a configuration
depends on the depth and weighted relative volume (configurational entropy) of
its potential energy basin.

Since lowest free energy corresponds to lowest potential energy \emph{and} high
relative volume of the potential energy basin, we are often specifically
interested in zero-dimensional regions where the potential energy is lowest.
However, the volume of the potential energy basins corresponding to these
regions typically include portions of all of their higher dimensional ancestor
regions. These ancestor regions should therefore be found and explored. Similarly,
computing kinetics involves a comprehensive mapping of the topology of paths
between regions, where the paths could pass through other regions of various
effective dimensions. Although paths would be expected to favor low dimensional
regions since they have the lowest energy, these paths could be long, requiring
many energy ups and downs, as well as backtracking, which could cause more
direct paths to be favored that pass through higher dimensional, higher energy
regions.

EASAL (i) directly atlases and navigates the complex topology of small assembly
configuration spaces (defined earlier), crucial for understanding free-energy
landscapes and assembly kinetics; (ii) avoids multiple sampling of
configurational (boundary) regions, and minimizes rejected samples, both
crucial for efficient and accurate computation of configurational volume and
entropy and (iii) comes with rigorously provable efficiency, accuracy and
tradeoff guarantees (see Section \ref{sec:complexity}). To the best of our
knowledge, no other current software provides such functionality.

\subsection{Related Work}
\subsubsection{Related Work on Geometric Algorithms} 
A generalization of Problem (\cone, \ctwo) arises in the robotics motion planning
literature with exponential time algorithms to compute a roadmap (a version of
atlas) and paths in general semi-algebraic sets
\cite{bib:canny-roadmap,canny-alg,basu}, with probabilistic versions to improve
efficiency \cite{kavraki1,kavraki2}. For the Cartesian configuration space of
non intersecting spheres, Baryshnikov et al. and Kahle characterize the
complete homology \cite{Baryshnikov08022013,Kahle2011}, viable only for
relatively small point-sets or spheres, while more empirical computational
approaches for larger sets \cite{PhysRevE,Bubenik10statisticaltopology} come
without formal algorithmic guarantees. A geometric rigidity approach was
primarily used to characterize the graph of contacts of arbitrarily large
jammed sphere configurations in a bounded region
\cite{Kahle2012,Connelly:Jamming}.

Unlike these approaches, the goal of EASAL is to describe the configuration
space of Problem (\cone, \ctwo). In addition, EASAL is
deterministic and its efficiency follows from exploiting special properties of
those semi-algebraic sets that arise as configuration spaces of point-sets
constrained by distance intervals. 

\subsubsection{Related Work on Molecular and Materials Modeling}
\label{ssub:MolecularRelatedWork} 
The simplest form of supramolecular self assembly and hence the simplest
application of Problem (\cone, \ctwo) is site-specific docking. Computational
geometry, vision and image analysis have been used in site-specific docking
algorithms \cite{Bespamyatnikh,Choi2004,pmid1549581,Duhovny2002,pmid15980490}.
Unlike the more general goals of EASAL, the goal of these algorithms is to
simply find site-specific docking configurations with optimal binding affinity.
While this depends on equilibrium free energy, docking methods simply
evaluate an approximate free energy function.

On the other hand, prevailing methods for direct free energy computation - that
must incorporate both the depth and relative weighted volumes (entropy) of the
free energy basin - use highly general approaches such as Monte Carlo (MC) and
Molecular Dynamics (MD) simulation. They deal with a notoriously difficult
generalization of Problem (\cone, \ctwo) 
~\cite{kaku,Andricioaei_Karplus_2001,Hnizdo_Darian_Fedorowicz_Demchuk_Li_Singh_2007,Hnizdo_Tan_Killian_Gilson_2008,Hensen_Lange_Grubmuller_2010,Killian_Yundenfreund_Kravitz_Gilson_2007,Head_Given_Gilson_1997,GregoryS201199,doi:10.1021/jp2068123}.
Ergodicity of these methods is unproven for configuration spaces of high
geometric or topological complexity with low energy, low volume regions (low
effective dimension) separated by high energy barriers. Hence they require
unpredictably long trajectories starting from many different initial
configurations to locate such regions and compute their volumes accurately.

While these methods are applicable to a wide variety of molecular modeling
problems, they do not 
take advantage of the simpler inter-molecular
constraint structure of assembly (\cone, \ctwo) compared to,
say, the intra-molecular folding problem (see \cite{assembly:folding}):
active constraint graphs that arise in assembly (see \figref{fig:3-trees}) 
yield convexifiable configuration spaces whereas
the folding problem has additional `backbone' constraints that prevent
convexification. 
Therefore, even though the energy functions used by
MC and MD can differ in assembly and folding, 
these methods miss out on critical advantages by not explicitly
exploiting special geometric properties of small assembly configuration spaces. 
EASAL on the other hand exploits such geometric properties 
via Cayley convexification. 

We do not review the extensive literature on (ab-initio) simulation or other
decomposition-based methods that are required to tractably deal with large
assemblies.  For small cluster assemblies from spheres, i.e., $n=1$ and $k\le
18$, there exist a number of methods to compute free energy and configurational
entropy of subregions of the configuration space
\cite{Holmes-Cerfon2013,Arkus2009,Wales2010,Beltran-Villegas2011,Calvo2012,Khan2012,Hoy2012,Hoy2014}.
Working with traditional Cartesian configurations, they must deal with
subregions that are comparable in complexity to the entire Cartesian
configuration space of small molecules such as cyclo-octane
\cite{Martin2010,Jaillet2011,Porta2007}. With $n=3$, there are bounds for
approximate configurational entropy using robotics-based methods without
relying on MC or MD sampling~\cite{GregoryS201199}. For arbitrary $n$ and
starting from MC and MD samples, recent heuristic methods infer a topological
roadmap
\cite{Gfeller_DeLachapelle_DeLos_Rios_Caldarelli_Rao_2007,Varadhan_Kim_Krishnan_Manocha_2006,Lai_Su_Chen_Wang_2009,10.1371/journal.pcbi.1000415}
and use topology to guide dimensionality reduction
\cite{Yao_Sun_Huang_Bowman_Singh_Lesnick_Guibas_Pande_Carlsson_2009}. In
particular \cite{Holmes-Cerfon2013} formally showed that their (and EASAL's)
geometrization is physically realistic, but, they directly search for
hard-to-find zero dimensional active constraint regions by walking
one-dimensional boundary regions of the Cartesian configuration space. In
addition they compute one and two dimensional volume integrals. 

To the best of our knowledge these methods do not exploit key features of
assembly configuration spaces that are crucial for EASAL's efficiency and
provable guarantees. These include Thom-Whitney stratification, generic
rigidity properties, Cayley convexification, and recursively starting from the
higher-dimensional interior and locating easy-to-find boundary regions of
exactly one dimension less. Using these and adaptive Jacobian sampling
\cite{Ozkan2014Jacobian}, EASAL can rapidly find all generically
zero-dimensional regions and can be used to compute not only one and two, but
also higher dimensional volume integrals \cite{Ozkan2014MainEasal}, as well as
paths that pass through multiple regions of various dimensions. This is
important for free energy and kinetics computation.

\subsubsection{Recent Work Leveraging EASAL}
\label{related}
EASAL variants and traditional MC sampling of the assembly landscape of two
transmembrane helices have recently been compared from multiple perspectives in
order to leverage complementary strengths \cite{Ozkan2014MC}. In addition,
EASAL has been used to detect assembly-crucial inter-atomic interactions for
viral capsid self-assembly~\cite{Wu2012,Wu2014Virus} (applied to 3 viral
systems: Minute Virus of Mice (MVM), Adeno-Associated Virus (AAV), and
Bromo-Mosaic Virus (BMV)). This work exploited symmetries and utilized the
recursive decomposition of the large viral capsid assembly into an assembly
pathway of smaller assembly intermediates. Adapting EASAL to exploit symmetries
was the subject of \cite{sym8010005}.

Though the submitted implementation can handle only two point-sets as input
($k=2$), for greater than 2 point-sets, the extension of the EASAL algorithm
and implementation have been shown to be straightforward
\cite{Ozkan2011,Ozkan2014MainEasal}. When $n=1$, i.e., each point-set is an
identical singleton sphere, exploiting symmetries leads to simpler computation.
EASAL has been used to compute 2 and 3 dimensional configurational volume
integrals for 8 assembling spheres for the first time
\cite{Ozkan2014MainEasal}, relying on Cayley convexification. Building upon the
current software implementation of EASAL, an adaptive sampling algorithm
directly leads to accurate and efficient computations of configurational region
volume and path integrals \cite{Ozkan2014Jacobian}.

%% file: sections/Theory.tex
\section{The Theory Underlying EASAL}
\label{sec:theory}

\begin{figure}[htpb]
\centering
\subfigure[Active constraint graph used to label a 2D node in (b); 
$(a_1, b_1)$ is the sole active constraint edge.]
{\label{fig:2DToy}\includegraphics[width=0.2\linewidth]{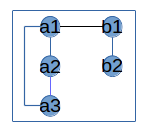}}
\subfigure[Stratification DAG of \exref{\toytwod}.]
{\label{fig:exampleStratification}
\includegraphics[width=0.7\linewidth]{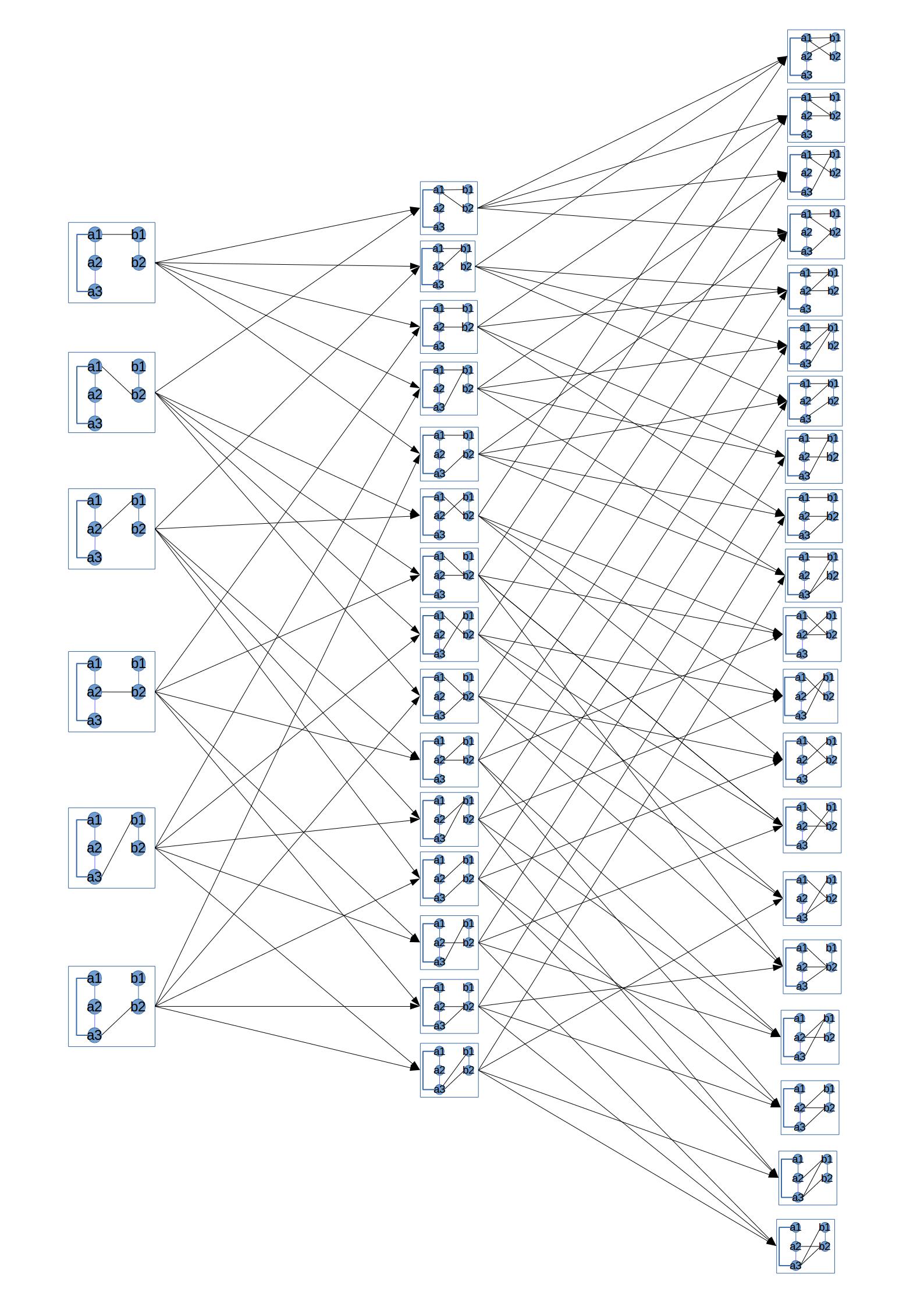}}
\caption{ Atlas (stratification) of the (toy-sized) configuration
space of \exref{\toytwod} of Section \ref{sec:toytwod}. (b) The nodes of the
DAG represent active constraint regions and DAG edges connect a region to a
boundary region, one dimension lower. Each node box displays the active
constraint graph of its corresponding region. The nodes in the leftmost column
represent 2D active constraint regions, i.e., they contain configurations with
two degrees of freedom. Adding an active constraint edge, yields 1D active
constraint regions (center column). Adding one more edge yields 0D regions,
each containing finitely many rigid configurations (rightmost column).}
\end{figure}

\begin{figure}[htpb]
\centering
\subfigure[Input to Problem (\cone, \ctwo).]{
\begin{overpic}[scale=.3,tics=10]{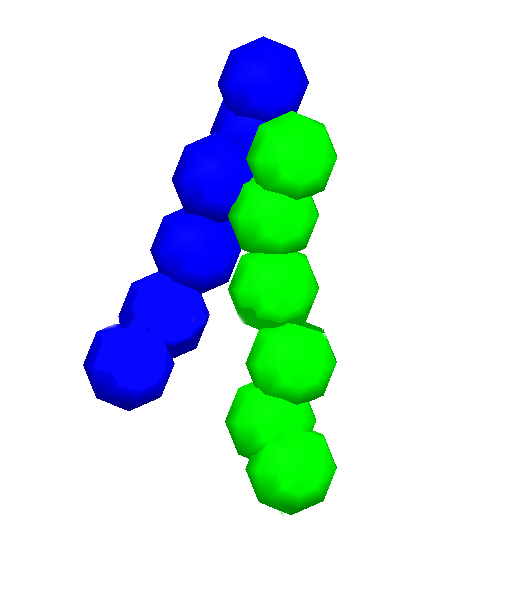}
     \put (20,60) {\color{blue}{$A$}}
     \put (53,60) {\color{green}{$B$}}
\end{overpic}
\label{fig:pctreeInput}
}
\subfigure[Active constraint regions in the atlas represented as nodes colored by their dimension, shown with their Cayley configuration spaces (see full caption below).]{
\begin{overpic}[scale=.3,tics=10]{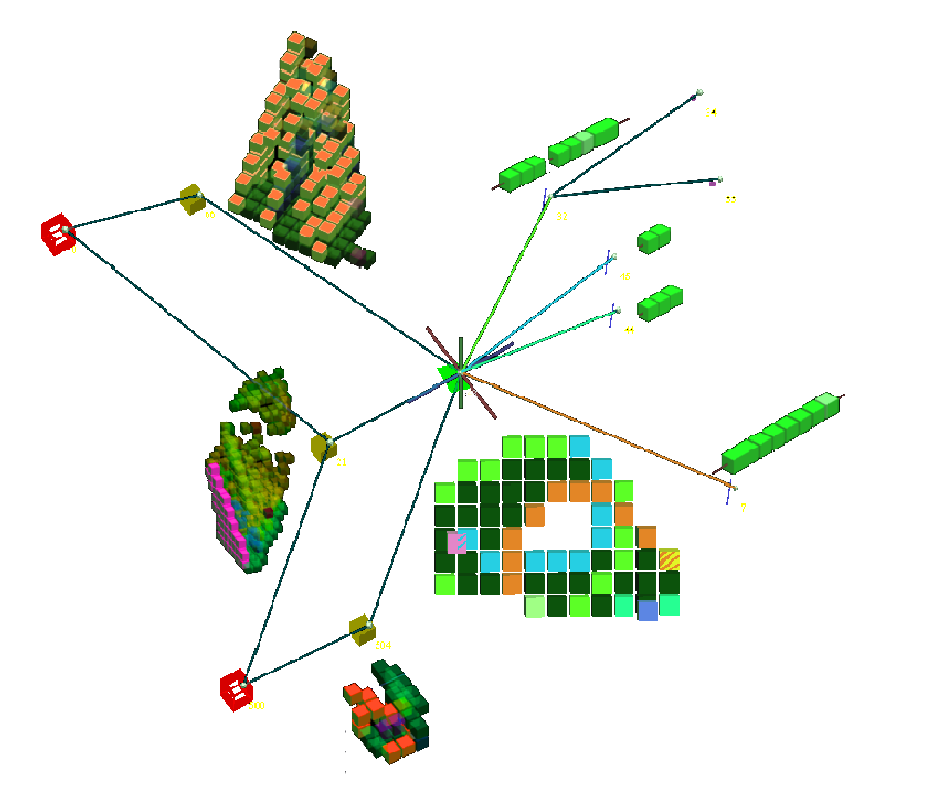}
     \put (50,75) {\color{grayoned}{1D Cayley}}
     \put (75,35) {\color{grayoned}{1D Cayley}}
     \put (75,28) {\color{grayoned}{1D node}}
     \put (76,20) {\color{green2d}{2D Cayley}}
     \put (76,15) {\color{green2d}{with hole}}
     \put (55,38) {\color{green2d}{2D node}}
     \put (08,80) {\color{yelloworange}{3D Cayley}}
     \put (48,05) {\color{yelloworange}{3D Cayley}}
     \put (42,15) {\color{yelloworange}{3D node}}
     \put (5,10) {\color{red}{4D node}}
\end{overpic}
\label{fig:pctreeSpace}
}
\subfigure[Active constraint regions in the atlas represented as nodes colored by their dimension, shown with their active constraint graphs.]{
\begin{overpic}[scale=.3,tics=10]{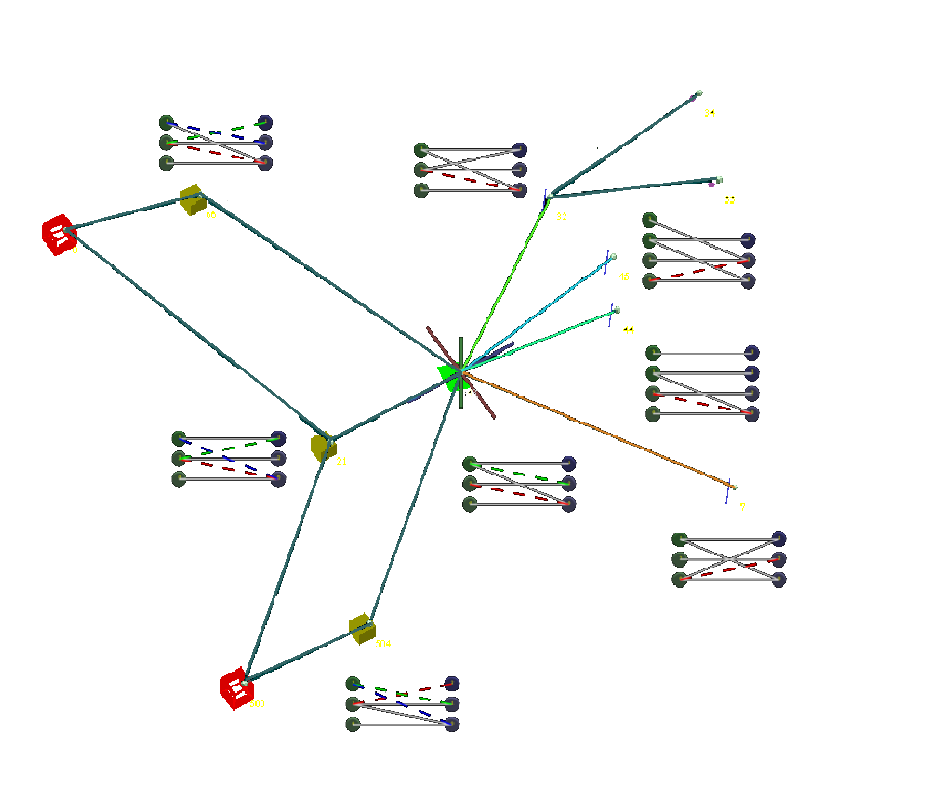}
     \put (69,18) {\color{grayoned}{5 active}}
     \put (65,14) {\color{grayoned}{constraints}}
     \put (75,28) {\color{grayoned}{1D node}}
     \put (48,27) {\color{green2d}{4 active}}
     \put (46,23) {\color{green2d}{constraints}}
     \put (40,37) {\color{green2d}{2D node}}
     \put (20,2) {\color{yelloworange}{3 active constraints}}
     \put (42,15) {\color{yelloworange}{3D node}}
     \put (5,10) {\color{red}{4D node}}
\end{overpic}
\label{fig:pctreeACG}
}
\subfigure[Active constraint regions in the atlas represented as nodes colored by their dimension, shown with Cartesian configuration sweep views(see full caption below).]{
\begin{overpic}[scale=.25,tics=10]{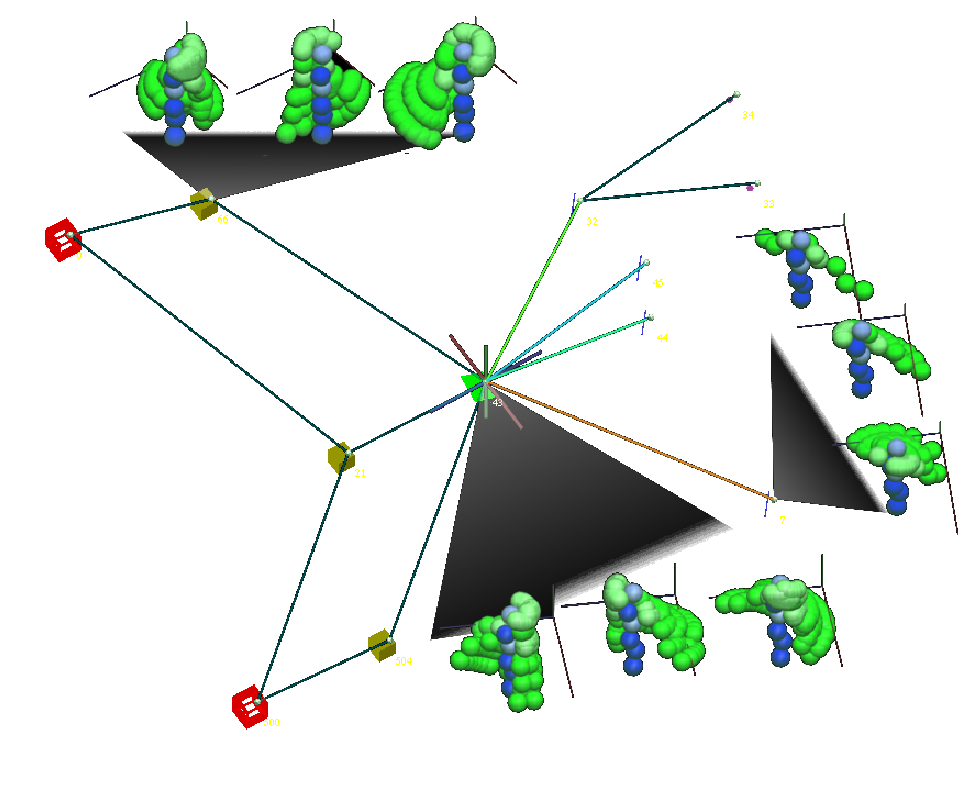}
     \put (15,85) {\color{yelloworange}{sweep views of different flips}}
     \put (75,25) {\color{grayoned}{1D node}}
     \put (40,40) {\color{green2d}{2D node}}
     \put (18,17) {\color{yelloworange}{3D node}}
     \put (0,10) {\color{red}{4D node}}
\end{overpic}
\label{fig:pctreeSweep}
}
\caption{\footnotesize 
(b), (c), and (d) show different views of a portion of the atlas centered on a
2D active constraint region.  (b) The grid of little cubes next to each node
delineates the Cayley configuration space of that region. Each little cube is a
Cayley point or a Cayley configuration.  Consider the 2D active constraint
region in the center.  This region has has no Cayley points in the middle (a
hole) since every realization of these Cayley points violates \cone. These
violations are caused by point pairs that are neither Cayley parameters nor
edges of the active constraint graph. Such hole regions typically also have a
convex Cayley parametrization.  The Cayley points highlighted with a different
color are points adjacent to their child (boundary) regions albeit using
different Cayley parameters.  (d) Each sweep view is the union of realizations,
one per Cartesian configuration in the corresponding node. Each sweep view
shows a different flip (defined in Section \ref{sec:realization}) of the Cayley
configuration space of the corresponding node. 
}
\label{fig:NestedRegions}
\end{figure}

The EASAL software is based on the theoretical concepts described in this section. 
We explain and illustrate EASAL's three strategies below. The reader will find
the video at \url{http://www.cise.ufl.edu/∼ sitharam/EASALvideo.mpg} useful to understand the following.

\subsection{Strategy 1: Atlasing and Stratification}
\label{sec:stratification}

EASAL's first strategy is to partition and stratify the Cartesian configuration
space into regions $R$ called the active constraint regions, each labeled by
its \emph{active constraint graph} (See \figref{fig:2DToy}).  Consider the set
of points participating in the active constraints that define $R$. Let $V_R$ be
any minimal superset of points that supports additional constraints, of type
\ctwo, to locally fix (generically rigidify) the two point-sets with respect to
each other.  Now, $V_R$ is taken to be the set of vertices of the active
constraint graph of $R$. An edge of the active constraint graph represents
either (i) one of the active constraints that define $R$ or (ii) a vertex pair
in $V_R$ that lies in the same point-set $A$ (or $B$) in Problem (\cone, \ctwo). Notice that building the active
constraint graph of $R$ reduces to picking a minimal graph isomorph from
\figref{fig:3-trees} containing the active constraints that define $R$.

The active constraint regions are organized as a partial order (directed
acyclic graph or DAG, see \figref{fig:exampleStratification} and
\figref{fig:pctreeACG}), that captures their dimensions and boundary
relationships. In particular, the active constraint graph of a region is a
subgraph of the active constraint graph of its boundary regions; and the
co-dimension of a region is generically the number of active constraint edges.
The analysis of the graph benefits from the following concepts of combinatorial
rigidity (we additionally refer the reader to \cite{CombinatorialRigidity}).

A \emph{linkage} is a graph, $G = (V, E)$, of vertices and edges, with an
assignment of lengths, $\gamma: E \rightarrow \mathbb{R}$, for each edge. A
(Euclidean) \emph{realization} of a linkage in $\mathbb{R}^3$ is an assignment
of points in $\mathbb{R}^3$ to vertices (factoring out the three rotations and
three translations of SE(3)) such that the Euclidean distance between pairs of
points are the given edge lengths $\gamma$. A realization is said to be
\emph{rigid} if there is no other realization in its neighborhood that has the
same edge lengths. A graph is said to be rigid if a generic linkage realization
of the graph is rigid. Otherwise, the graph is said to be flexible (not rigid).
A rigid linkage generically has finitely many realizations. A graph is said to
be \emph{minimally rigid}, \emph{well constrained} or \emph{isostatic} if it is
rigid and the removal of any edge causes it to be flexible. When the
realization is rigid, all non-edges have locally fixed lengths and are said to
be locally \emph{implied} or \emph{dependent}.  If the graph $G$ arises as an
active constraint graph for Problem (\cone, \ctwo) with the active constraint
edges being assigned length intervals, we obtain an \emph{active constraint
linkage}. In this paper we treat active constraint linkages just like linkages
while analyzing generic rigidity properties.

\begin{figure}[htpb]
\centering
      \subfigure[20 atom input]{
         \label{fig:3Input}
         \includegraphics[scale=0.2]{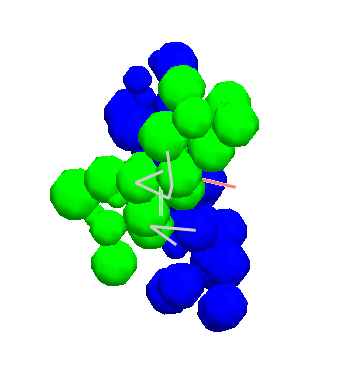}
         }
	\subfigure[Realization sweep view]{
	\label{fig:RealizationSweepView}
	\begin{overpic}[scale=.2,tics=10]{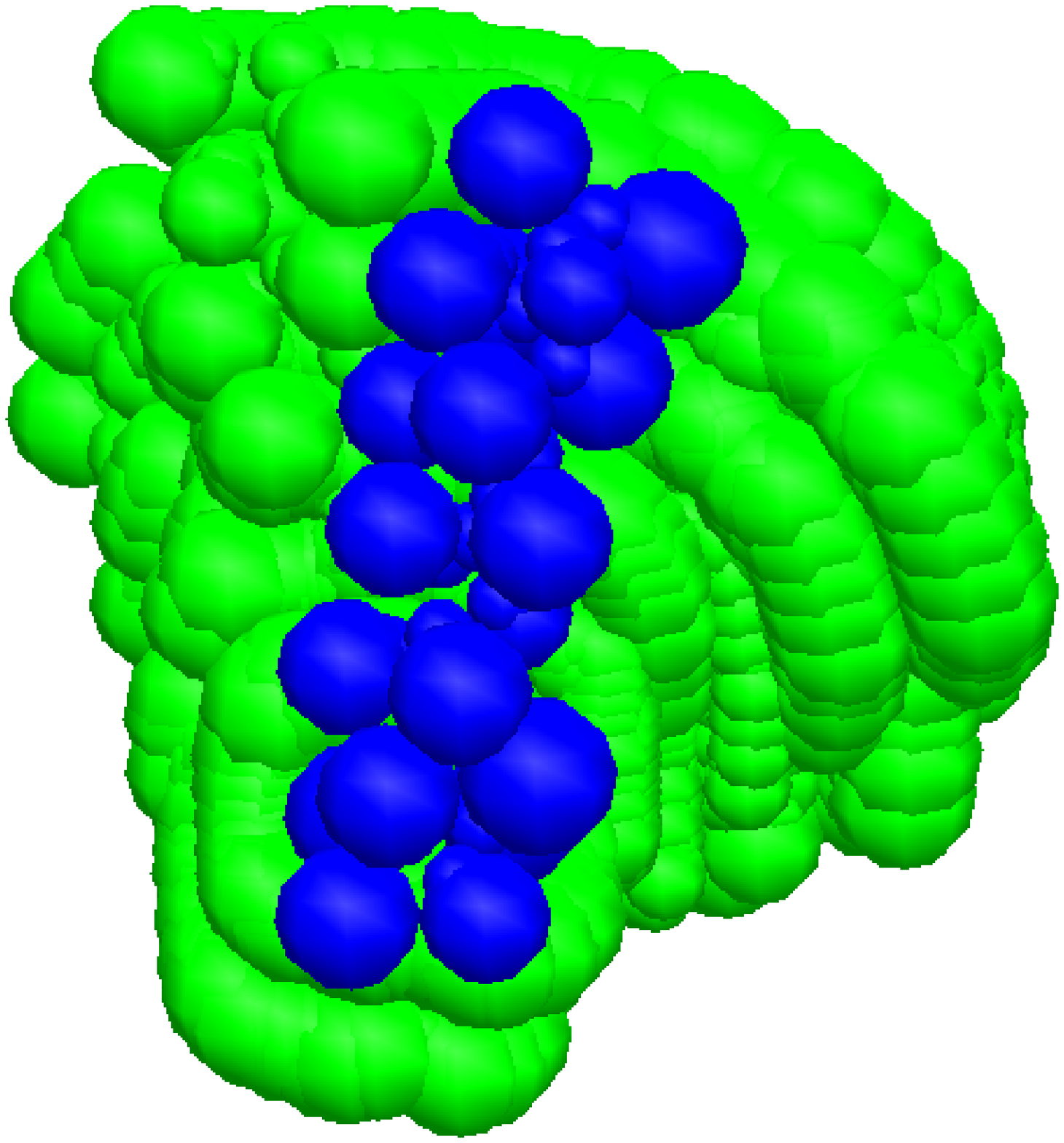}
		\linethickness{3pt}
		\put(56,76){\color{black}\vector(1,1){18}}
		\put(36,73){\color{black}\vector(1,3){8}}
		\put (20,100) {\color{black}{$T(B)$ sweep}}
		\put (73,95) {\color{black}{$A$ fixed}}
	\end{overpic}
        \includegraphics[scale=0.2]{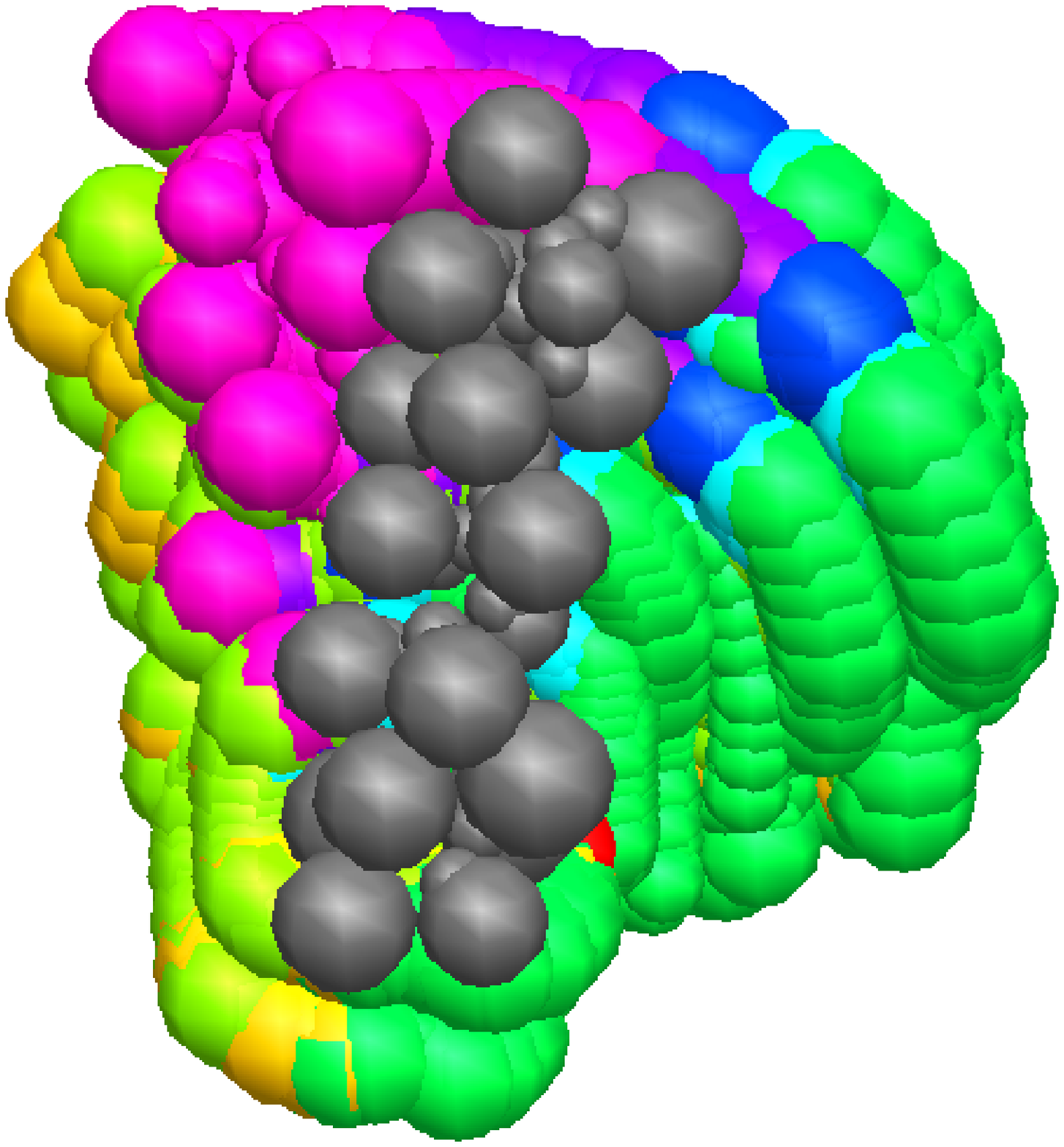}
        }
	\phantom{\includegraphics[scale=0.2]{\fig/Figure3Input.png}}
	\subfigure[Cayley configuration space view]{
           \includegraphics[scale=0.35]{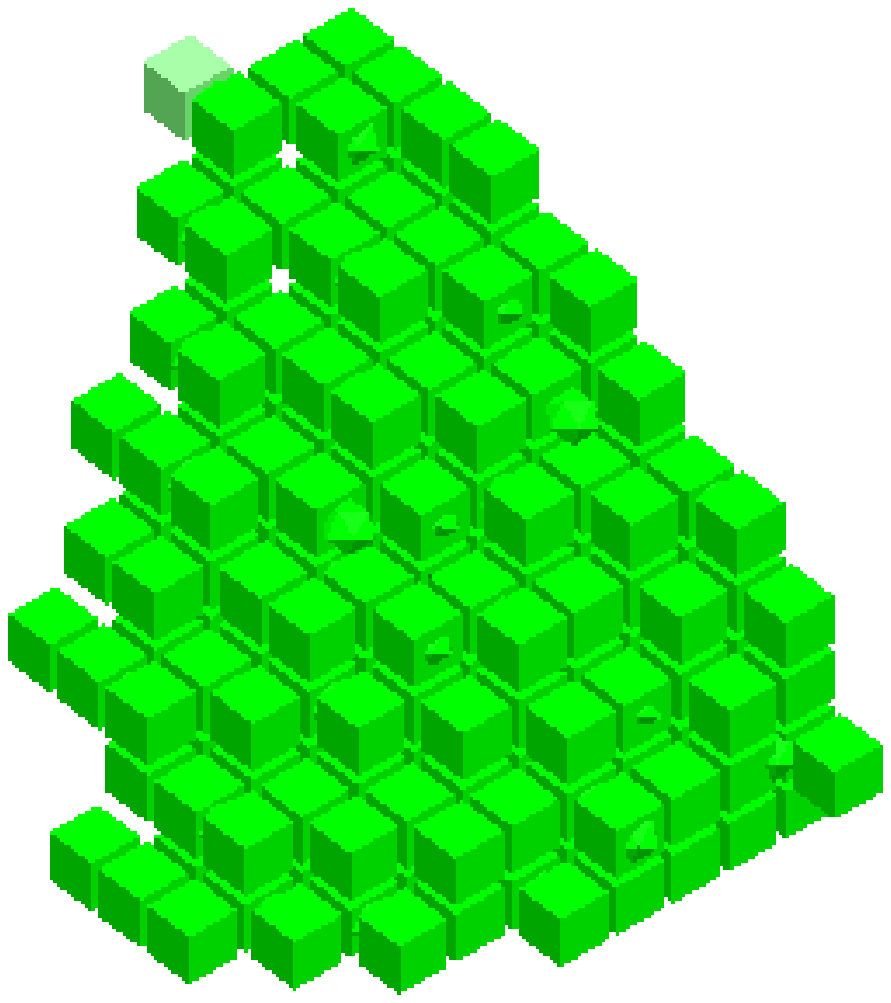}
           \includegraphics[scale=0.35]{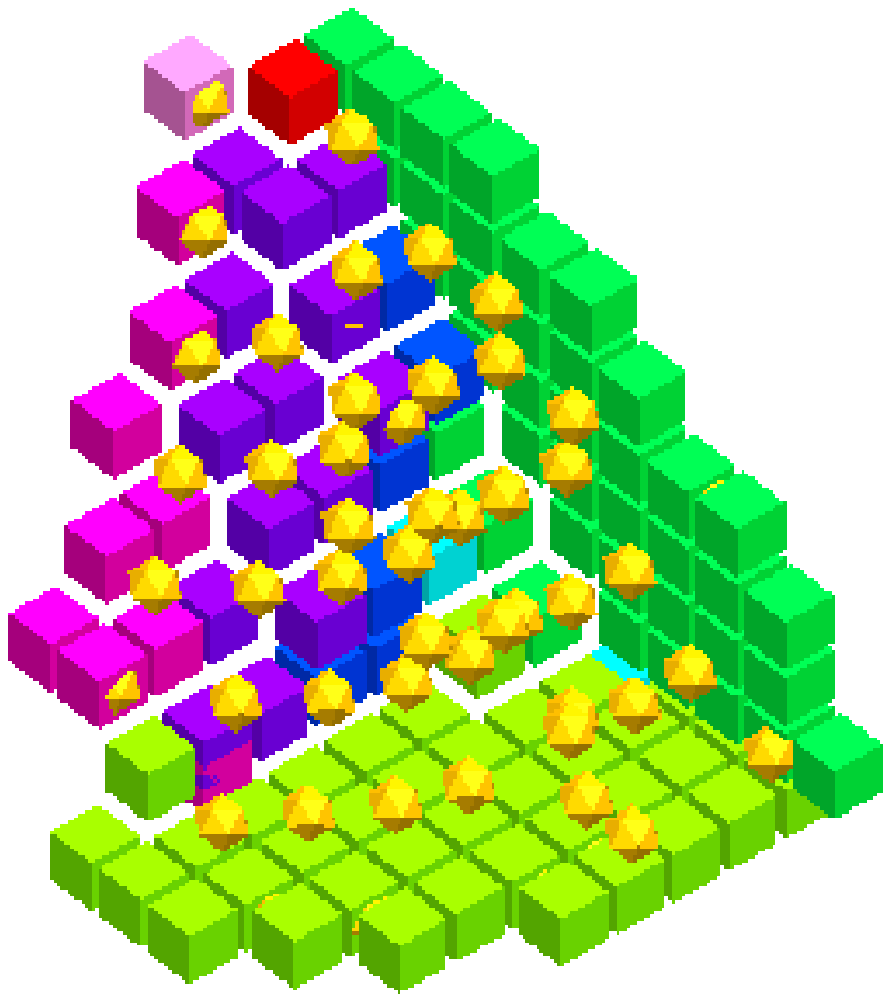}
           }
\caption{ Realization sweep view and Cayley configuration view of \exref{\bighelix}.
(b) Sweep of realizations $T(B)$ for fixed $A$ in a 3D active constraint
region. (b,\IR) Same view with $T(B)$ color coded to show
realizations adjacent to lower dimensional boundary regions where a new
constraint becomes active.
(c) Each cube represents one Cayley configuration with
at least one realization. 
(c,\IR) Only those Cayley points adjacent to child boundary regions are color coded as in (b, \IR),
except for the yellow ones, shown as icosahedra, which are placed there as
witness points (see Sections \ref{sec:exploration}, \ref{sec:aposteriori} and \ref{sec:raytracing}) 
by parent regions, since this region is a boundary of those parent regions.
}
\label{fig:boundarySweeps}
\end{figure}

The degrees of freedom (dof) of a graph (linkage) is the minimum number of
edges whose addition, generically, makes it rigid. Thus, the number of degrees of freedom is
the same as the generic (effective) dimension of the realization space of a
(active constraint) linkage of the graph. In $\mathbb{R}^3$, Maxwell's theorem
\cite{maxwell} states that rigidity of a graph $G=(V,E)$, implies that $|E| \ge
3|V| - 6$ (in $\mathbb{R}^2$, $|E| \ge 2|V| - 3$). If the edges are
independent, this ensures minimal rigidity.

{\bf Note} (Genericity Assumption): When $k=2$, the effective dimension of an
active constraint region plus the number of active constraints is always 6,
i.e., the number of active constraints generically determines the co-dimension
of the region. This is because, in Problem (\cone, \ctwo), generically, implied
non-edges are not active constraints, i.e., the active constraint edges are not
implied by (dependent on) the rest of the active constraint graph. Inactive
constraints (implied or not) do not restrict the dimension of active constraint
regions. For the special case of Problem (\cone, \ctwo), in which sets $A$ and
$B$ are centers of non intersecting spheres of generic distinct radii, these
assumptions are an unproven conjecture, for which counterexamples haven't been
encountered. When the radii are all the same, simple counterexamples exist
where implied non-edges are active constraints.

Employing these concepts, EASAL is able to use a classical notion called the
Thom-Whitney stratification \cite{Kuo} of (effective) dimensional regions of a
semi-algebraic set to stratify the configuration space atlas. In the atlas, DAG
edges between two nodes indicate a boundary relationship: a lower dimensional
child region is the boundary of a parent region one dimension higher (one fewer
active constraint). Thus, the atlas is organized into strata, one for each
(effective) dimension, and DAG edges exist only between adjacent strata. In
Section \ref{sec:exploration}, we describe in detail the algorithm used for
atlasing and stratification of the configuration space.

\subsubsection{\toytwod}
\label{sec:toytwod}
Consider Problem (\cone, \ctwo) in $\mathbb{R}^2$ with two point-sets $A$ and
$B$; $A$ contains three points - $a_1$, $a_2$, and $a_3$ and $B$ contains two
points - $b_1$ and $b_2$. The ambient space is SE(2) of dimension 3. A complete
stratification of the realization space is shown in
\figref{fig:exampleStratification}. The three strata are organized as a DAG,
with nodes representing active constraint regions and labeled by their
corresponding active constraint graphs.  The vertices in the active constraint
graph are points participating in the active constraints that define $R$. The
edges are of two types, (i) between points in the same point-set and (ii) the
active constraints, between points in different point-sets.

All regions in the leftmost column
consist of configurations with two degrees of freedom and are called 2D nodes.
Adding an extra active constraint to any of these nodes yields 1D nodes in the
center column. By adding an extra active constraint to the nodes in the center
column, we get the 0D nodes, shown in the rightmost column, each containing
finitely many rigid configurations. A DAG edge represents a boundary
relationship of the child region to a parent interior region one dimension
higher.

\subsection{Strategy 2: Recursive Search from Interior to Lowest Dimensional Boundary}
\label{sec:recursiveBoundarySearch}
To construct the atlas, EASAL's second strategy is to recursively, using depth
first search, start from the interior of an active constraint region and always
locate boundaries or child regions of strictly one dimension less. The
boundary or descendant regions of an active constraint region consist of
configurations where new constraints become active and lead to the discovery of
children active constraint regions. \figref{fig:boundarySweeps} shows the
boundary regions in the Cayley and Cartesian configuration spaces for a typical
3D active constraint region in a toy-sized atlas.

In particular, a boundary region with one additional active constraint
corresponds to 1 dimension less than the interior or parent region. Since
EASAL only looks for boundaries one dimension less at every stage (boundary
detection is explained in detail in Section \ref{sec:boundarydetection}), it
has a higher chance of success than looking for the lowest dimensional active
constraint regions directly (0D regions contain realizations of rigid active
constraint linkages, that are sought as low energy configurations in the
context of molecular and materials assembly).

Moreover, generically, if there is a region with the active
constraint set $H \cup \{a\} \cup \{b\}$, then the region with active
constraint set $H$ has at least two boundary or child regions, one with active
constraint set $H \cup \{a\}$ and another with active constraint set $H \cup
\{b\}$ as the active constraints. Both of these are parents of the region with
active constraint set $H \cup \{a\} \cup \{b\}$.

Because of this, when a new region is found, all its ancestor regions can be
discovered. So, even if a ``small'' (hard-to-find) region is missed at some
stage, if any of its descendants are found at a later stage, say via a larger
(easy-to-find) sibling, the originally missed region is discovered.

\subsection{Strategy 3: Cayley Convexification for Efficient Search and Realization} 
\label{sec:convexification} 

Locating a boundary region satisfying an additional active constraint is,
off-hand, challenging due to the disconnectedness and complexity of Cartesian
active constraint regions. To address this challenge, EASAL uses a theoretical
framework developed in \cite{SiGa:2010}. EASAL efficiently maps (many to one) a
$d$-dimensional active constraint region, to a convex region of $\mathbb{R}^d$ called
the Cayley configuration space. Convexity allows for efficient sampling and
search for boundaries. In addition, it is efficient to compute the inverse map
from each Cayley configuration to its finitely many corresponding Cartesian
realizations or configurations. We describe this strategy in more detail
below.


A \emph{complete 3-tree} is any graph obtained by starting with a triangle and
adding a new vertex adjacent to the vertices of a triangle in the current
graph. Alternatively, this amounts to successively pasting a complete graph on
4 vertices (a \emph{tetrahedron}) onto a triangle in the current graph. This
yields a natural ordering of vertices in a 3-tree (we drop `complete' when the
context is clear). A 3-tree has $3|V| -6 $ edges and hence, a 3-tree linkage
is minimally rigid in $\mathbb{R}^3$. That is, a 3-tree generically has
finitely many realizations, and removing any edge gives a flexible
\emph{partial 3-tree}. 

One way to represent the realization space of a flexible partial 3-tree linkage
is by choosing non-edges (called Cayley parameters) that complete it to a
3-tree. Then, given a partial 3-tree linkage and length values for the chosen Cayley
parameters there are only finitely many realizations for the resulting rigid
3-tree linkage. Since finitely many Cartesian realizations correspond to a
single Cayley configuration (tuple of Cayley parameter values), the Cayley
parametrization is a many to one map from the Cartesian realization space to
the Cayley configuration space. The inverse map can be computed easily by
solving three quadratics at a time as explained in Section
\ref{sec:algRealization}. Therefore, if the Cayley configuration space were
convex, it, and thereby the Cartesian realization space, can be efficiently
sampled.

Theorem \ref{thm} below asserts that the length tuples of non-edge Cayley
parameters $F$ (that complete a partial 3-tree into a 3-tree) form a convex
set. Given a linkage with edges $H$ of length $l_H$ a \emph{chart} for this 
linkage is defined by choosing a non-edge set $F$ with lengths $l_F$ such that
the linkage with edge set $H \cup F$, and edge lengths $l_H$ and $l_F$ is realizable.
Formally, the chart is the set $\{l_F: (H\cup F, l_H, l_F)$ is
realizable in $\mathbb{R}^3\}$, denoted  $\Phi_F(H\cup F, l_H)$.

\begin{theorem}{(\cite{SiGa:2010} Any partial 3-tree yields an exact convex chart)}
\label{thm} 
If an active constraint graph $G_H = (V, H)$  of an active constraint region $R$ is a partial
3-tree then, by adding edge set $F$ to give a complete 3-tree $G = (V, E = F
\cup H)$, we obtain an exact convex chart $\Phi_F (G, H, l_H)$ for $R$, in the
parameters $F$.  The exact convex chart $\Phi_F (G, H, l_H)$ has a linear
number of boundaries in $|G|$ defined by quadratic or linear polynomial
inequalities. If we fix the parameters in $F$ in sequence, their explicit
bounds can be computed in quadratic time in $|G|$.
\end{theorem}

As explained in \cite{SiGa:2010}, the theorem still holds when $H$ is an active
constraint linkage i.e., when $l_H$ is a set of intervals rather than a set of
fixed lengths. Besides proving Theorem \ref{thm} \cite{SiGa:2010} shows the
existence of convex Cayley configuration spaces for a much larger class of
graphs (beyond the scope of this paper).

Furthermore, as elaborated in \cite{Ozkan2014MainEasal}, for active constraint
graphs arising between $k$ point-sets, \emph{generalized 3-trees} yield convex
configuration spaces.  This is because each point-set represents a unique
realization of their underlying complete graph. A generalized 3-tree is defined
by construction similar to a 3-tree. However, during the construction, assume 3
or more vertices in the already constructed graph $G$ belong to the same
point-set say $A$ of Problem (\cone, \ctwo). Now, if a new vertex $v$ is constructed with edges to the
vertices of a triangle $T$ in $G$, then the $m \le 3$ vertices in $A \cap T$
can be replaced by any other $m$ distinct vertices in $A$ to which $v$ is
adjacent.  Moreover, generalized 3-trees, just like 3-trees, have an underlying
sequence of tetrahedra, and are rigid with finitely many realizations.  Going
forward, we simply refer to generalized (partial) 3-trees as (partial) 3-trees. 

\begin{figure}
\centering
\includegraphics[scale=0.25] {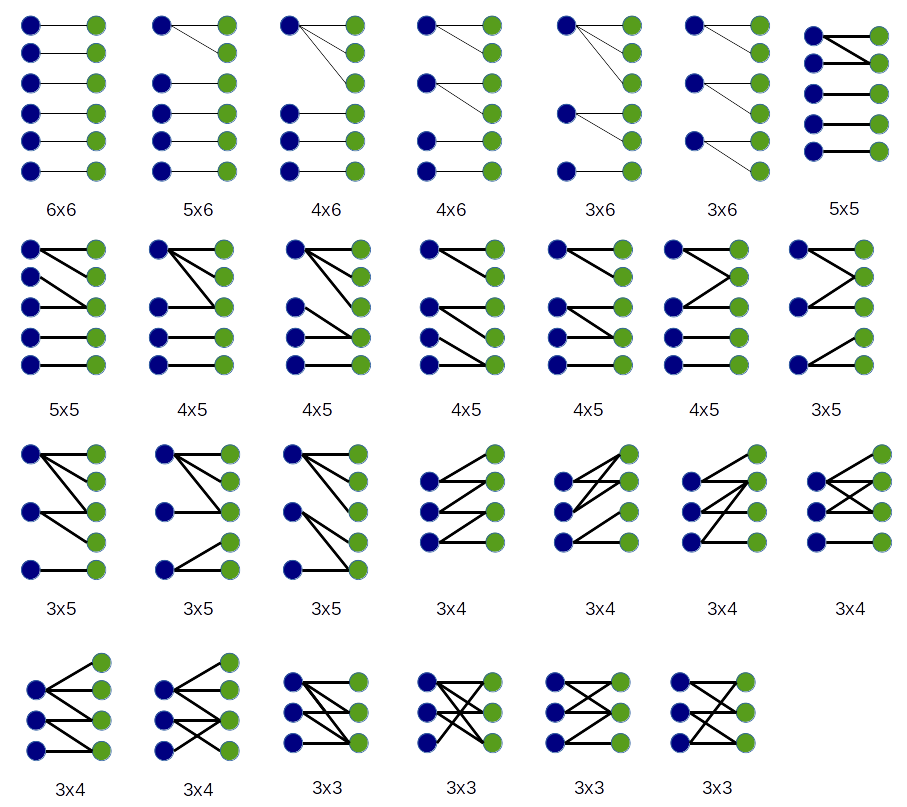}
\caption{ In each graph above, the vertices of the same color represent points
in the same point-set $A$ (or $B$) in Problem (\cone, \ctwo) and form a complete graph (whose edges are not shown).
Edges between vertices indicate point pairs whose distance is in the preferred
interval, i.e., the constraint is active. For $k=2$, all active constraint
graphs are isomorphic to subgraphs of the ones shown. The graphs above are
rigid and correspond to generically rigid 0-dimensional active constraint
regions. The label $m_1 \times m_2$ below each active constraint graph
indicates that $m_1$ points in the first point-set and $m_2$ points in the
second point-set participate in the active constraints.} 
\label{fig:3-trees}
\end{figure}

The quadratic and linear polynomials defined in Theorem \ref{thm} arise from
simple edge-length (metric) relationships within all triangles and tetrahedra
and are called \emph{tetrahedral inequalities}, and the explicit bounds
mentioned in the theorem are called \emph{tetrahedral bounds}. EASAL leverages
this efficient computation of the convex bounds enhanced by the Theorem 5.1.3
in \cite{ugandhar}, described in Section \ref{sec:tetrahedralbounds}.  It turns
out that, for small $k$, almost all active constraint graphs arising from
Problem (\cone, \ctwo) are partial 3-trees and thus their regions have a convex
Cayley parametrization. Specifically (see \figref{fig:3-trees}), all the active
constraint graphs with 1, 2 and 3 active constraints (5D, 4D and 3D atlas
regions) are partial 3-trees. $86\%$ of active constraint graphs with 4 active
constraints (2D atlas regions) and $70\%$ of active constraint graphs with 5
active constraints (1D atlas regions) are partial 3-trees. Since, regions with 6
active constraints (0D atlas regions) have finite realization spaces, Cayley
parametrization is irrelevant. Section \ref{sec:raytracing} describes how we
find realizations when the active constraint graph is not a partial-3-tree.

Although most active constraint graphs have convex Cayley configuration spaces,
the feasible region is a non-convex subset created by cutting out a region
defined by other constraints of type \cone. Each such constraint is between a
pair of points, one from each point-set, that is neither an active constraint
nor a Cayley parameter in the active constraint graph. However, the regions
that are cut out typically have a (potentially different) convex Cayley
parametrization. This can be seen in \figref{fig:pctreeSpace} where the Cayley
configuration space of the node in the center has a hole cut out because of
constraint violations by point pairs that are neither Cayley parameters nor
edges in the active constraint graph.

\subsubsection{\toytwod\ contd}
Here, the active constraint graph shown in \figref{fig:2DToy} is used to
illustrate Cayley convexification. Since that example is in $\mathbb{R}^2$,
2-trees serve the purpose of 3-trees used in EASAL \cite{SiGa:2010}. A
\emph{complete 2-tree} is any graph obtained by starting with an edge and
successively pasting a triangle onto an edge in the current graph. A
\emph{partial 2-tree} is any subgraph of complete 2-tree.

\begin{figure}[htpb]
\begin{center}
\includegraphics[width=.6\linewidth]{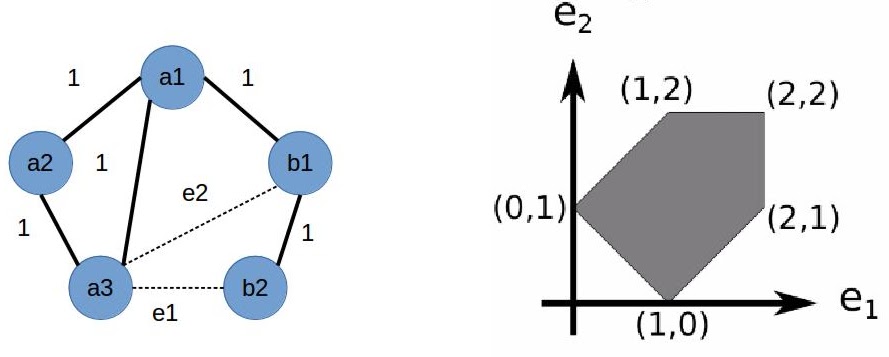}
\end{center}
\caption{\exref{\toytwod} viewed as
(\IL) a linkage in $\mathbb{R}^2$ (see text for description). (\IR)
The 2D convex Cayley configuration space for the linkage and the chosen Cayley
parameters $e1$ and $e2$. The shaded area delineates the realizable lengths of
$e1$ and $e2$.} 
\label{fig:2DCayley}
\end{figure}

Consider the partial 2-tree linkage shown in \figref{fig:2DCayley} (left).  To
represent the configuration space of this flexible linkage, we add the
non-edges $e1$ and $e2$, shown with dotted lines, to complete the 2-tree. This
not only makes the linkage rigid, but its realization is easy by a
straightforward ruler and compass construction, solving two quadratics at a
time. The non-edges $e1$ and $e2$ are the Cayley parameters and correspond to
independent flexes.  \figref{fig:2DCayley} (right) shows the convex Cayley configuration
space corresponding to this linkage.

If the edges in the graph in \figref{fig:2DCayley} (left) were assigned length
intervals instead of fixed lengths, yielding an active constraint linkage, the
resulting configuration space would continue to be convex, but would be 7
dimensional. However, when these intervals are relatively small in comparison
to the edge lengths, the Cayley configuration space remains effectively 2
dimensional.

\subsubsection{Realization: Computing Cartesian Configurations from a Cayley Configuration}
\label{sec:realization} 
The addition of the Cayley parameter non-edges to the active constraint graph
yields a complete 3-tree. This reduces computation of the Cartesian
realizations of a Cayley configuration (a tuple of Cayley parameter length
values) to realizing a complete 3-tree linkage. Realizing a complete 3-tree
linkage with $i$ tetrahedra reduces to placing $i$ new points one at a time
using 3 distance constraints between a new point and 3 already placed points.
For each new point we solve the quadratic system for intersecting 3 spheres
resulting in two possible placements of the new point. This yields $2^i$
possible realizations of the Cayley configuration. A \emph{flip} associated
with the Cayley configuration space consists of Cartesian realizations of all
Cayley configurations restricted to one of these $2^i$ placements
\cite{Ozkan2014MainEasal}. 


%% file: sections/AlgorithmicIdeas.tex
\section{Algorithmic Ideas and Implementation}
\label{algorithms}
This section discusses the key algorithmic ideas implemented in EASAL. EASAL
starts by generating all possible active constraint graphs with 1 or 2
(depending on user input) active constraints yielding 5D or 4D regions
(represented as root nodes) in the atlas and then successively samples them.
The main algorithm, ALGORITHM \ref{alg:sampleAtlasNode} merges the three
strategies described in the previous section.

\begin{algorithm} [htbp]
 \SetKwInOut{Input}{input}\SetKwInOut{Output}{output}

 {\bf sampleAtlasNode}\\
 \Input{atlasNode: node}
 \Output{Complete sampling of the atlasNode and all its children}
 \BlankLine

	$H$ = node.activeConstraints\\
	$G_H$ = node.activeConstraintGraph\\
	\If{ $G_H$ is minimally rigid}
		{stop;	}
	$F$ = complete3Tree($G_H$)\\
	
	$C$ = computeConvexChart($G_H$, $F$)\\

	\For{ each cayleyPoint $p$ within convexChart $C$ }
	{
		$R$ = computeRealizations($p$)\\

		\For{ each realization $r$ in $R$}
		{
			\If{!aPosterioriConstraintViolated($r$)}
			{
				\If{ isBoundaryPoint($r$) \&\& hasNewActiveConstraint($r$, $G_H$) }
				{
					$e$ = newActiveConstraint($r$, $G_H$);\\
					$G'$ := $G_{H \cup \{e\}}$ ;\\
					\If{ $G'$ is not already present in the current atlas}
					{
						childNode = new atlasNode($G'$)\\
						childNode.insertWitness($p$);\\
						sampleAtlasNode(childNode);\\
					} \Else{
						childNode = findNode($G'$);
					}
					node.setChildNode(childNode);
				} 
			}
		}
	}
	\caption{High level EASAL pseudocode}
\label{alg:sampleAtlasNode}
\end{algorithm}

ALGORITHM \ref{alg:sampleAtlasNode} proceeds as follows. It (i) recursively
(by depth first search) generates the atlas by discovering active constraint
regions of decreasing dimension; (ii) uses Cayley convexification of the region
to efficiently compute bounds for Cayley parameters a priori (before
realization), and samples Cayley configurations in this convex region; (iii)
detects boundary regions of 1 dimension less a posteriori (after realization)
i.e., when a new constraint becomes active, and efficiently finds the (finitely
many) Cartesian realizations of the Cayley configuration samples. We describe
each of these aspects of the algorithm in Sections \ref{sec:exploration},
\ref{sec:tetrahedralbounds} and \ref{sec:boundarydetection} respectively.

\subsection{Atlasing and Stratification}
\label{sec:exploration}
EASAL stores and labels regions of the Cartesian configuration space as an
atlas as described in Section \ref{sec:stratification}. The regions of the
atlas are stored as nodes of a directed acyclic graph, whose edges represent
boundary relationships. Each region of the atlas is an active constraint region
associated with a unique active constraint graph $G_H$, where $H$ is the set of
active constraints (see Algorithm 1). 

\begin{figure}
\centering
\includegraphics[width=\textwidth] {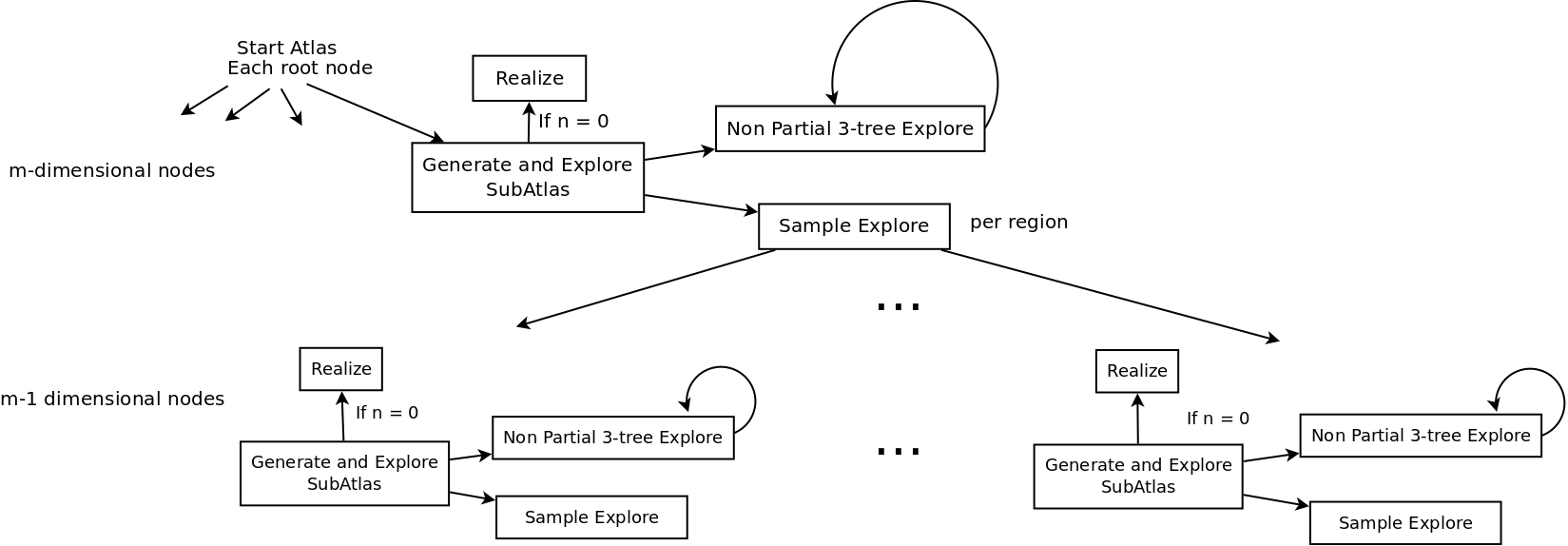}
\caption{A high level flowchart of the algorithm for generating and exploring the atlas}
\label{fig:algorithm}
\end{figure}

The exploration of the atlas is done by the recursive \textbf{sampleAtlasNode}
algorithm using one of the generated atlas root nodes as input. Using depth
first search, this algorithm samples the atlas node and all its descendants.
\figref{fig:algorithm} gives an overview of the algorithm.

\noindent\textbf{Base case of recursion:} If active constraint graph $G_H$ of
the node is minimally rigid i.e., the active constraint region is 0D, then
there is only 1 Cayley configuration (with finitely many Cartesian
realizations).  We have no more sampling to do, hence return.

\noindent\textbf{The recursion step:} If $G_H$ is not minimally rigid, EASAL
applies the \textbf{complete3Tree} algorithm of in Section
\ref{sec:tetrahedralbounds} to find a set of parameters $F$ to form a 3-tree.
This leverages the convex parametrization theory~\cite{SiGa:2010} of Section
\ref{sec:convexification} and ensures that a linkage with edge set $H \cup F$
is minimally rigid and easily realizable.

Next EASAL finds the convex chart for the parameters $F$ via the
\textbf{computeConvexChart} algorithm. This algorithm leverages Theorem
\ref{thm} enhanced by the theory presented in \cite{ugandhar}. This algorithm,
detects the tetrahedral bounds and samples uniformly within this region using a
user specified step size. Detection of the tetrahedral bounds is explained in
more detail in Section \ref{sec:tetrahedralbounds}.

Next we compute the Cartesian realization space of the convex chart using the
\textbf{computeRealization} algorithm (described in Section
\ref{sec:algRealization}). This uses two nested for loops. The outer loop runs
for each Cayley point $p$ in the convex chart and computes the realizations for
each of these points as described in Section \ref{sec:realization}. The inner
loop runs for each realization $r$ of the point $p$ and detects whether some
Cayley points violate constraints between pairs that do not form an edge of
active constraint graphs.  {\sl This is the crucial test that indicates that a
new constraint has become active.} The Cayley point whose realization caused a
child boundary region to be found at a parent is called a \emph{witness} point,
since it witnesses the boundary, and is placed in the child boundary region
clearly labeled as a witness point coming from each parent region (see also
Figure
\figref{fig:boundarySweeps} and Section \ref{sec:raytracing}). We perform the
\textbf{aPosterioriConstraintViolated} check (described in Section
\ref{sec:boundarydetection}) to discover a boundary region. For every new
region discovered in this manner, we sample the region recursively with the
\textbf{sampleAtlasNode} algorithm.

\subsection{Cayley Convexification and A Priori Computation of Bounds}
\label{sec:tetrahedralbounds}
According to the theory of convex Cayley parametrization in Section
\ref{sec:convexification}, if the active constraint graph of an active
constraint region is a partial 3-tree, choosing non-edges that complete the
partial 3-tree into a complete 3-tree as Cayley parameters always yields a
convex Cayley space. In other words, the active constraint linkage has a convex
Cayley configuration space if it is a partial 3-tree.  Computing the bounds of
this convex region ensures that sampling stays in the feasible region and
minimizes discarded samples.

The first step is thus to find the set of Cayley parameters that complete a
partial 3-tree. This is done by the \textbf{complete3Tree} algorithm. The
\textbf{complete3Tree} algorithm uses Theorem \ref{thm} of Section
\ref{sec:convexification}. It first creates a look-up table containing all
possible complete 3-trees. Given a graph $G_H$ as the input, we find a graph in
the look-up table so that $G_H$ is a proper subgraph of either the graph or one
of its isomorphisms. The set of edges by which $G_H$ differs from the graph
found in the look-up table is returned as $F$. $F$ is the set of Cayley
parameters.

Finding bounds for each Cayley parameter (bounds on edge lengths for $F$) has
two cases:
\begin{itemize}
\item[--] If there is only one Cayley parameter in a tetrahedron, the tentative
range of that parameter is computed by the intersection of tetrahedral
inequalities.

\item[--] If there is more than one unfixed Cayley parameter in a tetrahedron,
then the tentative ranges of a parameters are computed in a specific sequence
\cite{ugandhar}. The tentative range of a parameter in the sequence is computed
through tetrahedral inequalities using fixed values for the parameters
appearing earlier in the sequence. Since the range of the parameter is affected
by the previously fixed parameters, more precise range computation of the
unfixed parameter is required for every iteration/assignment of fixed
parameters.
\end{itemize}

The actual range for each parameter is obtained by taking the intersection of
the tentative range and the range of \ref{eqn:preferredConstraints}. The
order in which Cayley parameters are fixed have an effect
on the efficiency of the range computation~\cite{ugandhar}. We pick parameters
in the order that gives the best efficiency. Once we choose the parameters $F$
and the sequence, the explicit bounds can be computed in quadratic time in
$|G|$.  Once explicit bounds for each Cayley parameter have been found, we
populate this region by sampling it uniformly using a user specified step size. 

\subsection{Boundary Region Detection}
\label{sec:boundarydetection}
The boundary regions of an active constraint region caused by newly active
constraints can be detected only after Cartesian realizations are found using
the \textbf{computeRealization} algorithm (described later in this section).

If the newly active constraint occurs between a point pair that is a Cayley
parameter, then this is immediately detected at the start of sampling from the a
priori bounds computation of the convex Cayley region. In particular, if (i)
the actual range of a Cayley parameter $p$ for a region $r$ includes either the
lower or upper bound $\overline{p}$ of Problem (\cone, \ctwo)
and (ii) a Cayley point with $p = \overline{p}$
has a realization, then that Cayley point is on a boundary region of $r$.
Otherwise, if a newly active constraint occurs between a pair that is not a
Cayley parameter, then the corresponding boundary is detected as follows.

\subsubsection{A Posteriori Boundary or New Active Constraint Detection}
\label{sec:aposteriori}
A posteriori boundary detection involves checking for violation of constraints
corresponding to pairs that are neither edges nor Cayley parameters in the
active constraint graph. EASAL relies on Cayley parameter grid sampling to find
the child boundary regions of each active constraint region. However, boundary
detection is not guaranteed by Cayley parameter grid sampling alone, since the
sampling step size may be too large to identify a close-by point pair that
causes a newly active constraint. That is, the constraint violation could occur
between 2 feasible sample realizations or between a feasible and an infeasible
realization on the same flip in the sampling sequence. In the former case, the
missed boundary region is ``small.'' However, due to the precise structure of
Thom-Whitney stratification, it is detected if any of its descendants is found
via a larger sibling (as described in detail in Section
\ref{sec:recursiveBoundarySearch}). In the latter case, the newly active
constraint has been flagged but exploration (by way of binary search) is
required to find the exact Cayley parameter values at which new constraints
became active. The binary search is on the Cayley parameter value, with
direction determined by whether the realization is feasible or not.

In both cases, once a new active constraint $e$ is discovered, we add the new
constraint to $G_H$ and create an new active constraint graph $G' = G_{H \cup
\{e\}}$.  Notice that a boundary region could be detected via multiple
parents.
However, since regions have unique labels, namely the active constraint graphs,
no region is sampled more than once. If $G'$ has already been sampled, we just
add the node for $G'$ into the atlas, as a child of $G_H$. Otherwise, we create
a new atlas node with $G'$, sample it using the recursive
\textbf{sampleAtlasNode} algorithm and then add it as a child of $G_H$.
In both cases, the parent leaves one or more witness Cayley points in the 
child region (see Figure \ref{fig:boundarySweeps} and Sections
    \ref{sec:exploration} and \ref{sec:raytracing}).

\subsection{Cartesian Realization}
\label{sec:algRealization}
The \textbf{computeRealization} algorithm used to find realizations takes in an
active constraint region and its convex chart and generates all possible
Cartesian realizations. As stated earlier, each Cayley configuration can
potentially have many Cartesian realizations or flips. There are 2 cases
depending on whether the active constraint graph is a partial 3-tree or not.
Cartesian realization for partial 3-trees is straightforward as described in
Section \ref{sec:realization}. We describe the other case in detail next.

\subsubsection{Cartesian Realization for Non-partial 3-trees: Tracing Rays}
\label{sec:raytracing}
According to Section \ref{sec:convexification}, active constraint regions
without a partial 3-tree active constraint graph occur rarely.  To find tight
convex charts that closely approximate exact charts, we first drop constraints
one at a time, until the active constraint graph becomes a partial 3-tree. In
doing so, we end up in an ancestor region, with a partial 3-tree active
constraint graph and a convex Cayley parametrization. Note that since
non-partial 3-trees potentially arise only when we are exploring active
constraint regions with 4 or 5 active constraints (2D and 1D atlas nodes
respectively), it is always possible to drop one or two constraints to reach an
ancestor region which has a partial 3-tree active constraint graph. We do not
explore 0D regions. They consist of a single Cayley configuration with only
finitely many realizations, which are found when the region is found.

Once in the ancestor region, we trace along rays to populate the lower
dimensional region by searching in the ancestor region. For example, to find a
2D boundary region which does not have a partial 3-tree active constraint graph
or a convex parametrization, we drop one constraint. We then uniformly sample
the 3D region guaranteed to have a convex parametrization (setting the third
coordinate to zero). For each sample point, we traverse the third coordinate
using binary search (Section \ref{sec:aposteriori}). This generalizes to any
dimension and region in the sense that ray tracing is robust when searching for
and populating a region one dimension lower. By recursing on the thus populated
region, we find further lower dimensional regions.

\subsection{Complexity Analysis}
\label{sec:complexity}
The highest dimension of an active constraint region for $k=2$ is 6.  More
generally, for $k$ point-sets, the maximum dimension of a region is $6(k-1)$.
If $r$ regions of dimension $d$ have to be sampled, EASAL requires time linear
in $r$ and exponential in $d$.  Specifically, given a step size $t$ (a measure
of accuracy) as a fraction of the range for each Cayley parameter, the
complexity of exploring a region is $O((\frac{1}{t})^{6(k-1)})$. This indicates 
a tradeoff between complexity and accuracy \cite{Ozkan2011}. 

The complexity is also affected by $n$ the number of points in each point set.
This is due to a posteriori constraint checks which involve checking every point
pair (one from each point set) for violation of \cone. Thus, the complexity of 
exploring a region is $O((\frac{1}{t})^{6(k-1)} \times n^2)$.

If $r$ is the number of regions to explore, given as part of the input by 
specifying a set of active constraints of interest, the complexity of exploring 
all these regions is $O(r \times (\frac{1}{t})^{6(k-1)} \times n^2)$.
In the worst case, $r$, can be as large as $O(k^2 \cdot n^{12k})$. 
In this case, we cannot expect better efficiency, since the complexity cannot be 
less than the output size. Usually, $r$ is much smaller $O(k^2 \cdot n^{12k})$, since much 
fewer active constraint regions are generally specified as part of the input.

%% file: sections/Results.tex
\section{Results}
\label{sec:results}
In this section we briefly survey experimental results appearing in
\cite{Ozkan2014MainEasal,Ozkan2014MC,Wu2014Virus}, that illustrate some of EASAL's
capabilities. The main applications of EASAL are in
estimating free-energy, binding affinity, crucial interactions for assembly,
and kinetics for supramolecular self-assembly starting from rigid molecular
motifs e.g., helices, peptides, ligands etc.

\subsection{Atlasing and Paths} 
\label{sec:atlasandpath}
\begin{table}[h]
\resizebox{\columnwidth}{!}{%
\begin{tabular}{cC{4cm}cccc}\hline
		$n$ & Step size(as a fraction of the smallest radius) & Number of Regions & Number of samples & Good Samples & Time(in minutes)\\\hline\hline
	6 & 0.25  &26k &1.9 million & 1.3 million & 82\\\hline
	6 & 0.375 &23k & 617k & 379k & 23\\\hline
	6 & 0.5 & 19k& 289k & 172k & 11\\\hline
	20& 0.25  & 184k & 5.8 million& 716k & 335\\\hline
	$20^*$  & 0.25  & 206 & 63k & 22k & 2\\\hline
	$20^\dag$ & 0.25  & 3107 & 74k & 33k & 7\\\hline
\end{tabular}
\newline
}

\caption{Time on a standard laptop (see text) to stratify the configuration
space of pairwise constrained point-sets with the tolerance set to $(1.0-0.75)
\times$  sum of radii. The input point-set with $n=6$ is \exref{\toyhelix} and
the $n=20$ input is \exref{\bighelix}. Note that in $20^*$, only one 
5D and its children 4D regions are sampled and in $20^\dag$, only one 5D and its 
descendant 4D and 3D region are sampled.} 
\label{table:stratification}
\end{table}

In this section, we survey numerical results
from experiments in \cite{Ozkan2014MainEasal}, illustrating the performance of
EASAL in generating an atlas and computing paths for the configuration space of
two ($k=2$) input point-sets. The experiments were run on a machine with 
Intel(R) Core(TM) i7-7700 @ 3.60GHz CPU with 16GB of RAM. These results can be
reproduced by the reader using the accompanying EASAL software implementation
(see Section 2.4 of the User Guide for instructions).

The time required for generating the atlas is measured for a given accuracy of
coverage, measured in terms of the step size, and a given tolerance, which is
the width of the interval in \ctwo. Two different input point-sets 
(\exref{\toyhelix} and \exref{\bighelix}) are used as input. The results for 
the $n=6$ input
(\exref{\toyhelix}) show the time and number of
samples for generating the atlas of all possible combinations of active
constraint regions with one active constraint (5D atlas root nodes). The
results for $n=20$ input (\exref{\bighelix}) show the time and number of samples 
required to
generate the atlas for a typical randomly chosen 5D active constraint region and
all its children. Also note that in $20^*$, only one 5D 
and its children 4D regions are sampled, and in $20^\dag$, only one 5D and its 
descendant 4D and 3D regions are sampled. \exref{\bighelix} is challenging due 
to the number of ``pockets'' in the point-set structure leading to a highly 
intricate
topology of the configuration space with many effectively lower dimensional
regions. Table \ref{table:stratification} summarizes the results. These results
can be reproduced using the test driver submitted (see Section 2.4 of the user
guide).

\subsubsection{Finding Neighbor Regions}

\begin{figure}[htpb]
\centering
\begin{overpic}[scale=0.4,tics=10]{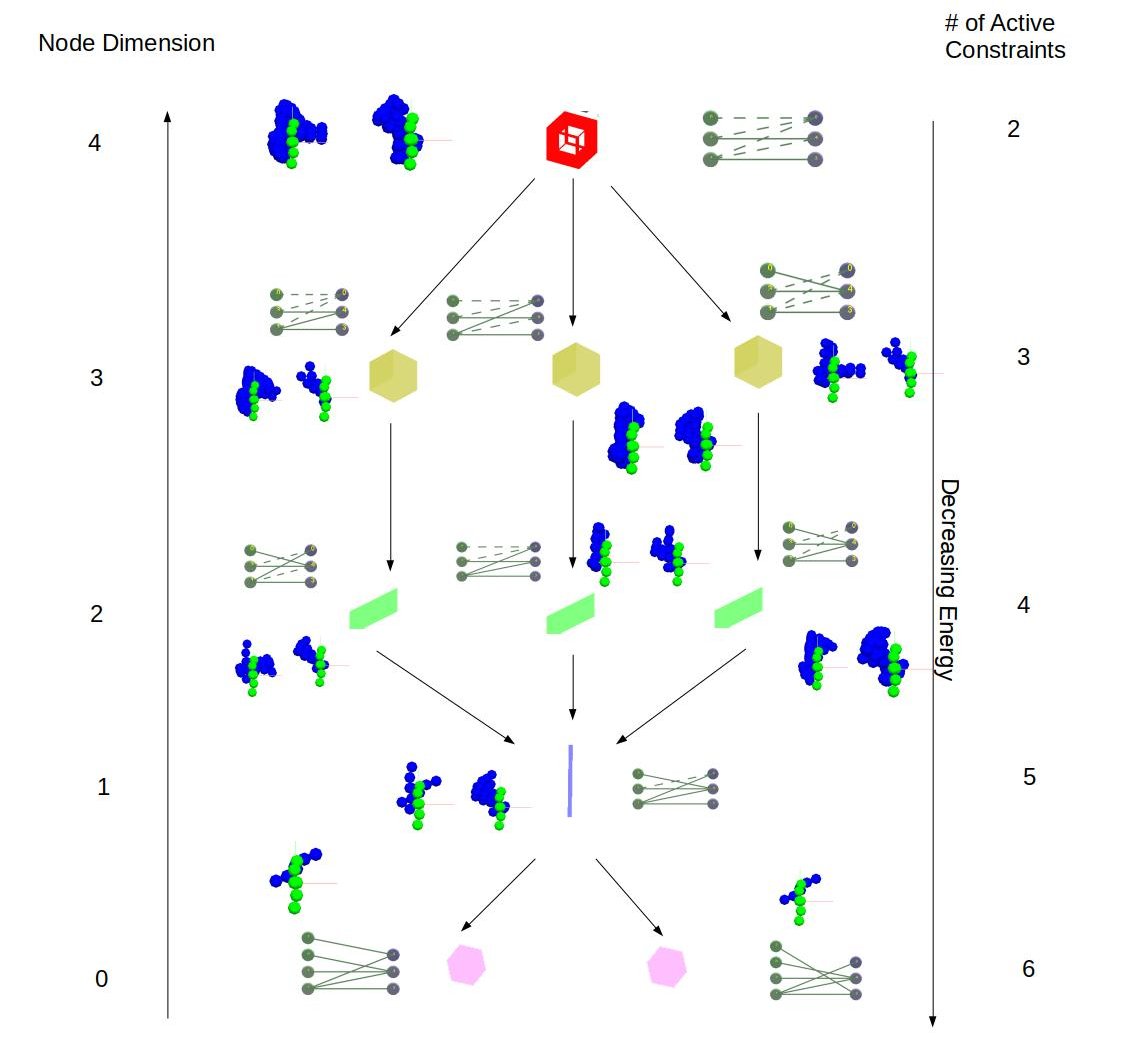}
     \put (28,85) {\color{blue}{Sweeps}}
     \put (48,85) {\color{red}{4D node}}
     \put (56,60) {\color{yelloworange}{3D node}}
     \put (55,37) {\color{green}{2D node}}
     \put (52,25) {\color{purple}{1D node}}
     \put (48,05) {\color{pink}{0D node}}
     \put (60,85) {\color{black}{Constraint Graph}}
\end{overpic}
\caption{
A portion of a toy-sized atlas. The ancestor and descendant regions, of
dimension four or less, of an active constraint region with 5 active
constraints (which is a 1D atlas region, shown here as a blue line). The pink
nodes represent its 0D child regions, the green nodes represent its 2D parent
regions, the beige nodes represent its 3D grandparent regions and the red node
represents its 4D ancestor region. Next to each node is shown its corresponding
active constraint graph and the sweep views of two flips.  The increasing
number of constraints reduces the potential energy of the assembly.
}
\label{boundingregions}
\end{figure}
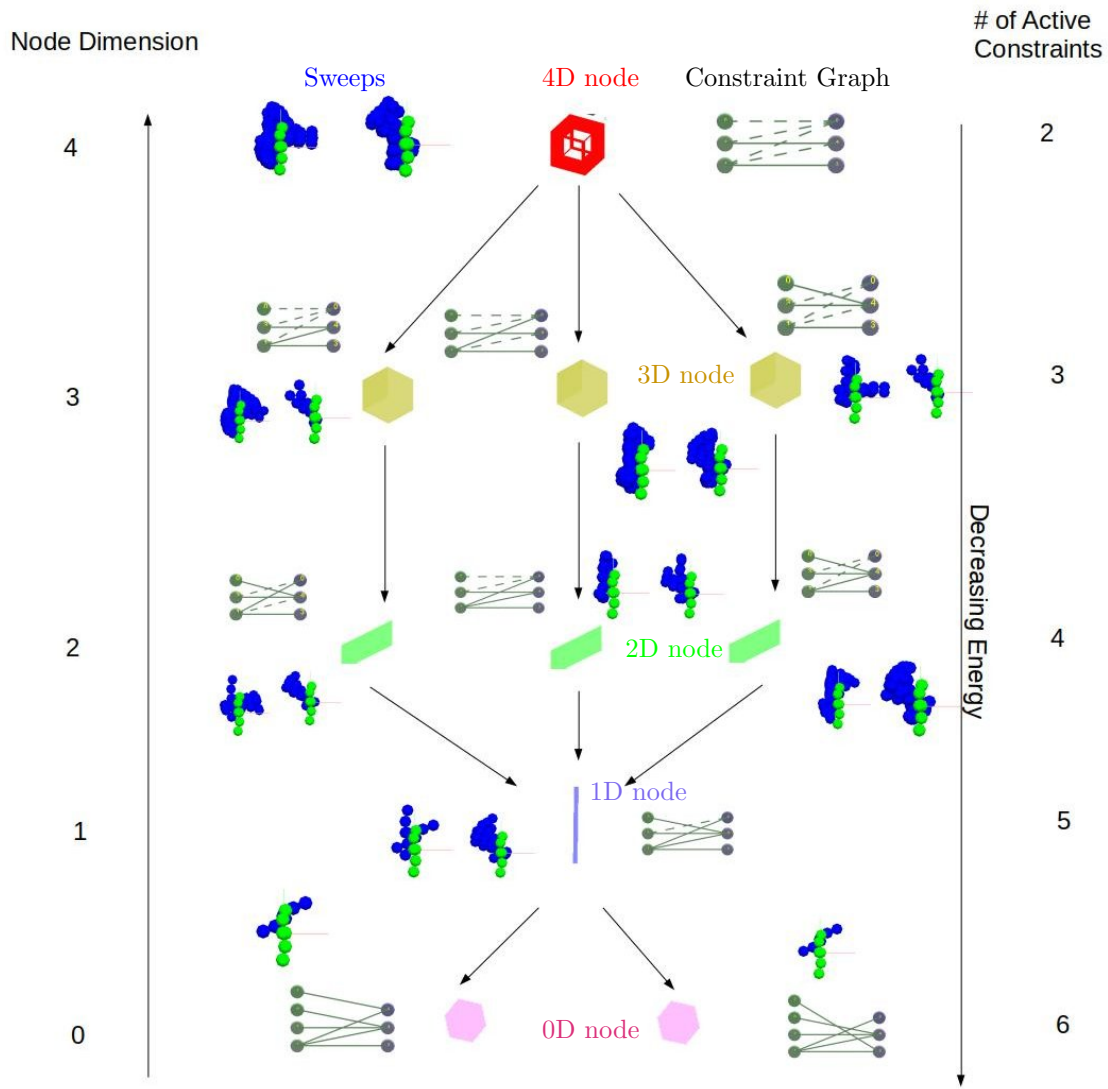

For any given active constraint region, one of EASAL's implemented
functionalities gives all of its neighbor regions.  A higher dimensional
neighbor (parent) region has one active constraint less and a lower dimensional
neighbor (child) region has one more active constraints. The atlas contains
information on child and parent regions of every active constraint region. 

If EASAL has been run using just the backend, the atlas information can be
accessed from the RoadMap.txt file in the data directory.  The neighbors are
listed as ``Nodes this node is connected to'' at the end of each node's
information.

In the optional GUI (not part of TOMS submission), the neighbors of a region
are listed in the Cayley space view.  The GUI contains a feature called `Tree',
which additionally shows all the ancestors and descendants of an active
constraint region.  \figref{boundingregions} shows the `Tree' feature being
used on an active constraint region having 5 active constraints. The figure
shows each ancestor and descendant node along with their active constraint
graphs and sweep views of Cartesian configurations in the region.

\subsubsection{Finding Paths between Active Constraint Regions}
The atlas output by EASAL can be used to generate all the paths between any two
active constraint regions along with their energies. Once the atlas has been
generated, finding paths is \emph{extremely fast} as we discuss below.

\begin{figure}[htpb]
\centering
\fbox{\includegraphics[scale=.15]{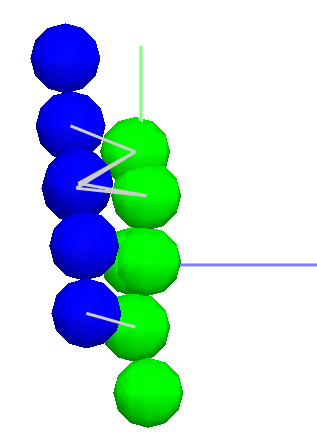}}
\includegraphics[scale=.4]{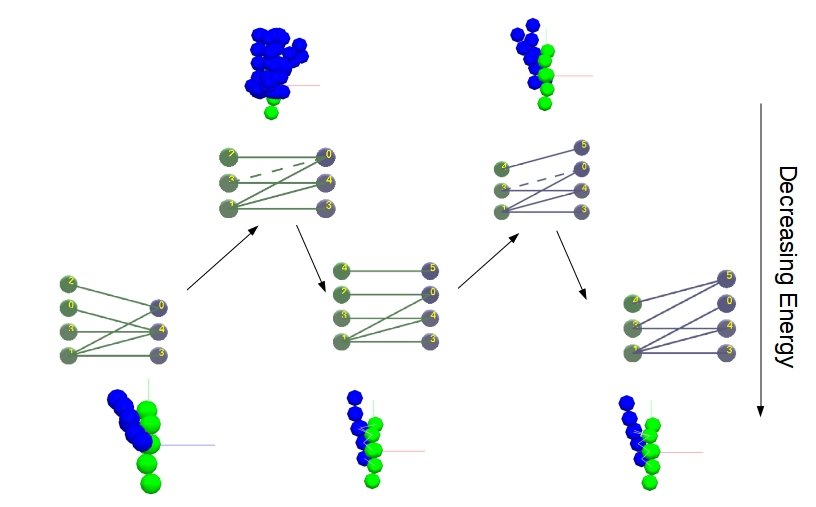}
\caption{A path in a toy-sized atlas. The path connects two active constraint
regions (left to right), each with 6 active constraints. The path traverses
regions with at most one less constraint. Each active constraint region is
labeled by its corresponding active constraint graph. The arrows form a path,
losing or gaining a new active constraint, from the source to the destination
active constraint regions.  The sweep view of feasible configurations of a
sample flip is shown next to each active constraint region. The left inset figure 
(\exref{\toyhelix}) shows the input molecules used for this experiment.}
\label{paths}
\end{figure}

Of particular interest is finding paths between two configurational regions
with zero degrees of freedom or with 6 active constraints. These are the 0D
nodes of the atlas with effectively rigid configurations. They find
paths in which the highest number of degree of freedom level is bounded. In
particular, paths through regions with 5 active
constraints with one step higher degree of freedom and one fewer constraint.
These regions represent a generic one degree of freedom motion path (see
\figref{paths}). 

\begin{table}[h]
\begin{tabular}{ccC{3cm}C{3cm}}\hline
$n$ & $r$ & Average length of shortest path &Average time (see text) to find shortest path\\\hline
6& 176 & 7 & 1.9 ms \\\hline
6& 145 & 6 & 2.2 ms \\\hline
20& 787 & 18 & 119ms\\\hline
\end{tabular}
\\\subcaption{\\The time on a standard laptop (see text), to find the shortest path
between two active constraint regions with 6 active constraints through $m$ 
other active constraint regions with 5 or 6 active constraints.}

\bigskip

\begin{tabular}{cccC{4cm}}\hline
$n$ & $r$ & $t$ & Average time (see text) to find number of paths\\\hline
\multirow{3}{*}{6}
				&176 & 2 & 2.02 s\\
				&176 & 4 & 4 s\\
				&176 & 8 & 6.04 s\\
				&176 & 10 & 8.08 s\\\hline
\multirow{3}{*}{20}
				& 787 & 2 & 6 min\\
				& 787 & 4 & 11.58 min\\
				& 787 & 8 & 18.04 min\\
				& 787 & 10 & 27.44 min\\\hline
\end{tabular}
\\\subcaption{\\The time on a standard laptop (see text), to find the number of
paths of length $t$, between all pairs of active constraint regions with
6 active constraints, in a toy atlas with $r$ active constraint regions with 6
active constraints.}
\caption{Finding paths between active constraint regions}
\label{table:paths}
\end{table}

This experiment was performed on two example point-sets with $n=6$ (\exref{\toyhelix})
and $n=20$ (\exref{\bighelix}). In the first experiment, the shortest
path between 100 randomly chosen pairs of active constraint regions with 6
active constraints are found. As shown in Table \ref {table:paths}, for the $n = 6$
example input, it took an average of 2 ms to find the shortest path, and the
average length of the shortest path was 6. For the $n=20$ example input, it
took an average of 119 ms to find the shortest path with the average length of
the shortest path being 18. These results can be reproduced using the test
driver (see Section 2.4 of the user guide)

In the second experiment, the number of paths of length $t$ in a toy
atlas between all pairs of active constraint regions are found. This toy atlas had $r$
active constraint regions with 6 active constraints. As shown in Table \ref
{table:paths}, for the example input with $n=6$, the number of paths
of length 10 were found in 8 seconds and the number of paths for the $n=20$ input in 27
minutes. These results can be reproduced using the test driver (see Section 2.4
of the user guide)

\subsection{Coverage and Sample Size Compared to MC}
In this section we sketch results from \cite{Ozkan2014MC} comparing EASAL and
its variants to the Metropolis Markov chain Monte Carlo (MC) algorithm for
sampling a portion of the landscape of two point-sets arising from protein
motifs (transmembrane helices, \exref{\bighelix}).  In that paper, the
effectiveness of EASAL in sampling crucial but narrow, low effective
dimensional regions is demonstrated by showing that EASAL provides similar
coverage as the traditional methods such as MC but with far fewer samples. For
determining coverage, it is sufficient to sample only the interior of an active
constraint region having 1 active constraint, without generating its children.

\begin{figure}[htpb]
\centering
\includegraphics[scale=0.2]{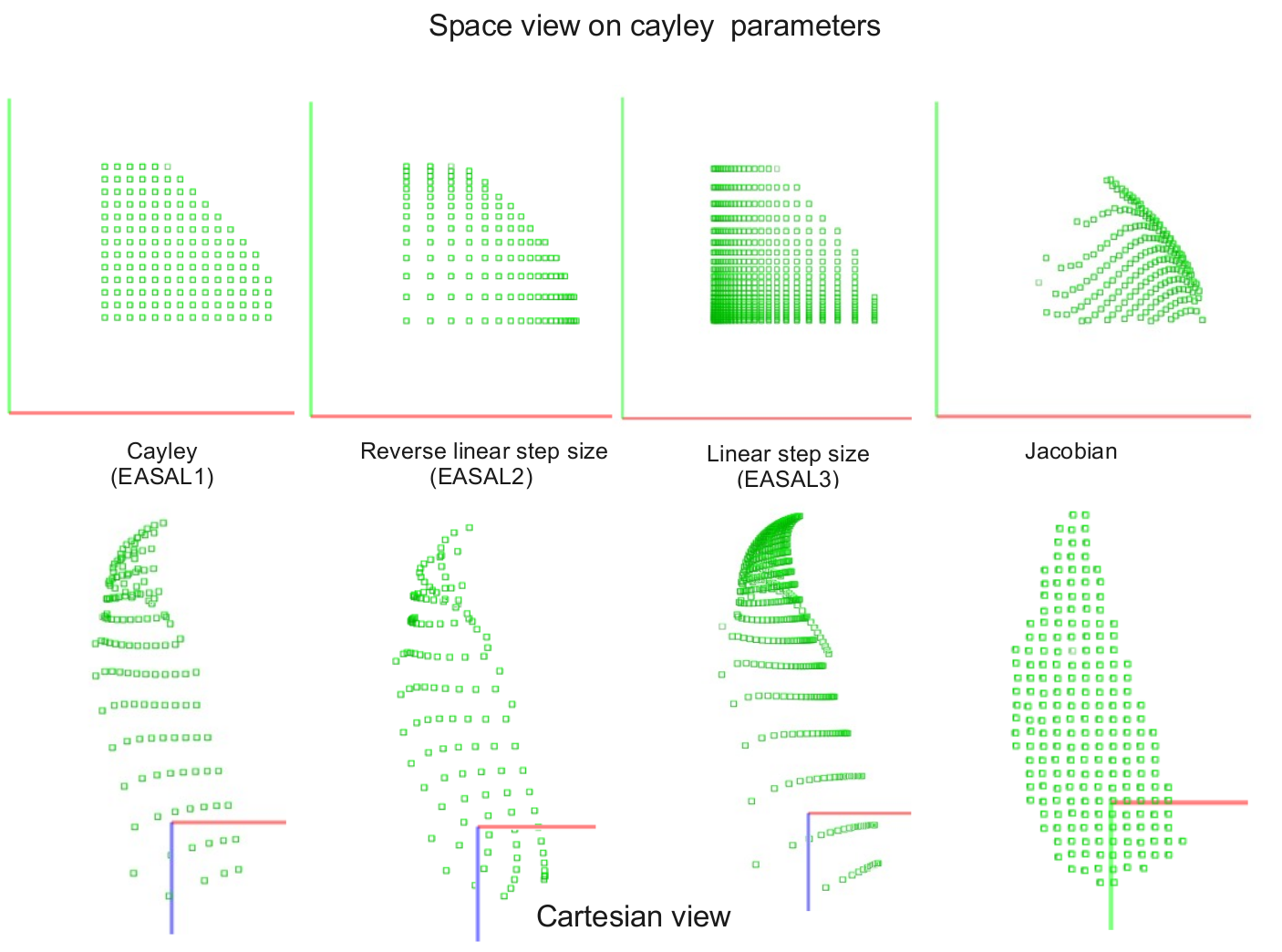}
\caption{Comparison of sampling in Cayley v/s Cartesian space in variants of
EASAL for a 2D active constraint region in the atlas for the example in
\figref{fig:3Input} \protect\cite{Ozkan2014MC}. The axes in the top
	figure are the two Cayley parameters. In the bottom figure, 
	the projection is on the $xy$ coordinates of the
centroid of the second point-set with the centroid of the first point-set fixed
at the origin.}
\label{fig:nonuniform}
\end{figure}

EASAL variants EASAL-1, EASAL-2, EASAL-3, and EASAL-Jacobian differ in their
sampling distributions in the Cayley space and by extension in the Cartesian
space.  EASAL-1 samples the Cayley space uniformly. Since energy is directly
related to distance, this does uniform sampling across energy levels.  This
however, skews the sampling in the Cartesian space. EASAL-2 uses a step size
inversely proportional to the Cayley parameter value. This samples more densely
in the interiors of the active constraint region and near tetrahedral bounds.
This is useful if we want to sample densely at places where degeneracies such
as flip intersections (so called conformational shifts and tunneling) are
likely to occur.  EASAL-3 uses a step size linearly proportional to the Cayley
parameter value.  This samples densely close to the boundaries. This is useful
if we want to sample densely at lower energy values.  EASAL-Jacobian uses a
sophisticated adaptive Cayley sampling method to force uniform sampling in the
Cartesian space. This is essential to compute volumes and thereby entropy and
free energy accurately. A comparison of how sampling in the Cayley space
relates to sampling in the Cartesian space, for these variants of EASAL, is
shown in \figref{fig:nonuniform}.

\begin{table}[h]
\resizebox{\columnwidth}{!}{%
\begin{tabular}{lcccccc}\hline
	sampling method & EASAL-1 &EASAL-2& EASAL-3& EASAL-Jacobian& MC &MultiGrid\\\hline
$\varepsilon$-coverage& $\lceil 0.97\rceil$ & $\lceil 1.14\rceil$& $\lceil 1.20\rceil$& $\lceil 0.66\rceil$ &$\lceil0.31\rceil$&N/A\\\hline
	Coverage percentage &92.06\% &92.42\% &74.08\% &99.53\% &99.96\%&N/A\\\hline
	Number of Samples & 100k & 40k & 30k & 1 million & 100 million&12 million\\\hline
	Ratio percentage & 3.56\%& 5.17\%& 2.97\% &3.45\% &1.29\%&N/A\\\hline
\end{tabular}
}
\caption{Comparison of EASAL variants with MC with respect to coverage and number
of samples for the two transmembrane helices shown in
\figref{fig:3Input} \protect\cite{Ozkan2014MC}. 
Here, $\varepsilon$ is computed as described in the text.}
\label{table:coverage}
\end{table}

\begin{figure}
\centering   
\subfigure[MC]{\label{fig:a}\includegraphics[scale=0.25]{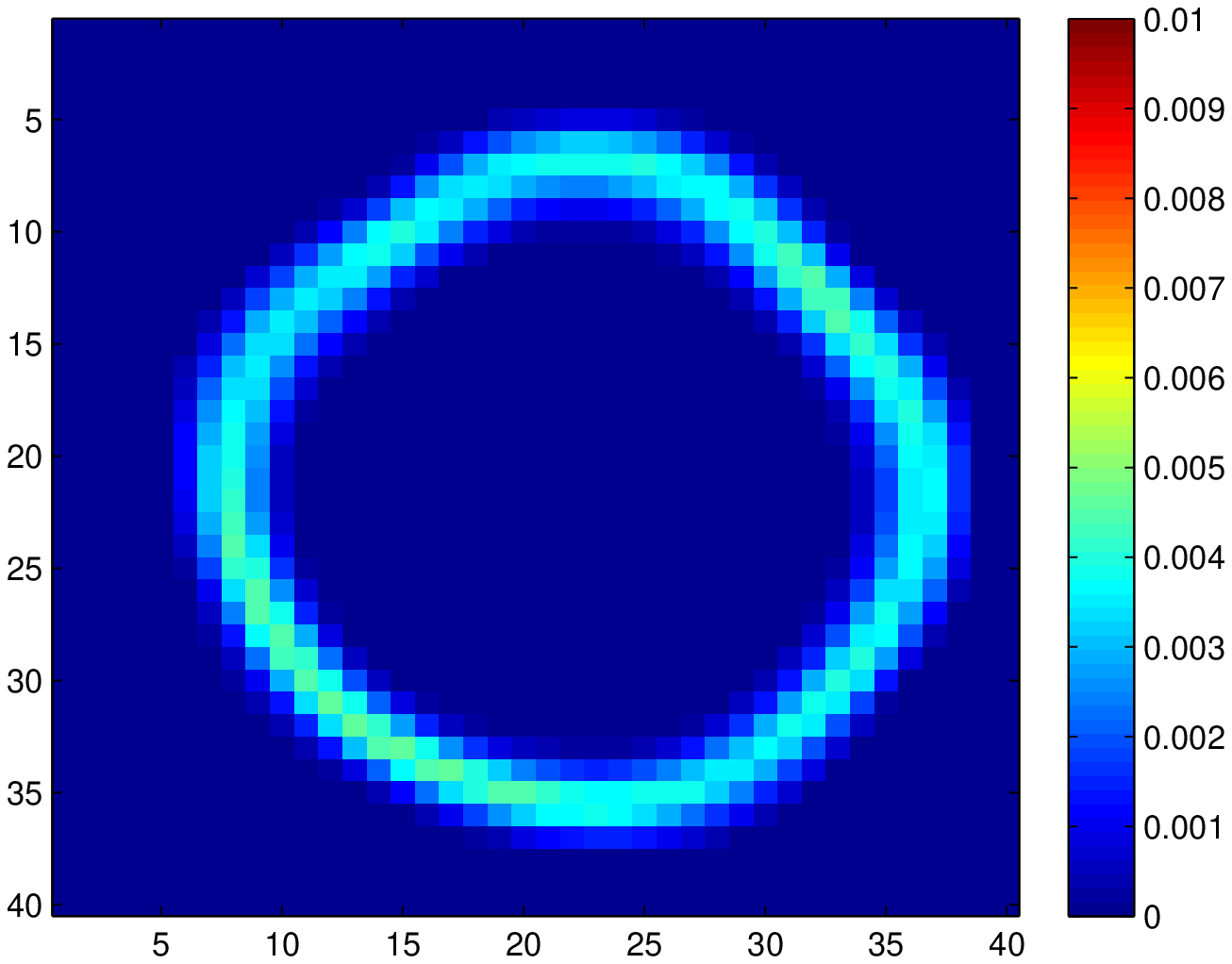}}
\subfigure[MultiGrid]{\label{fig:b}\includegraphics[scale=0.25]{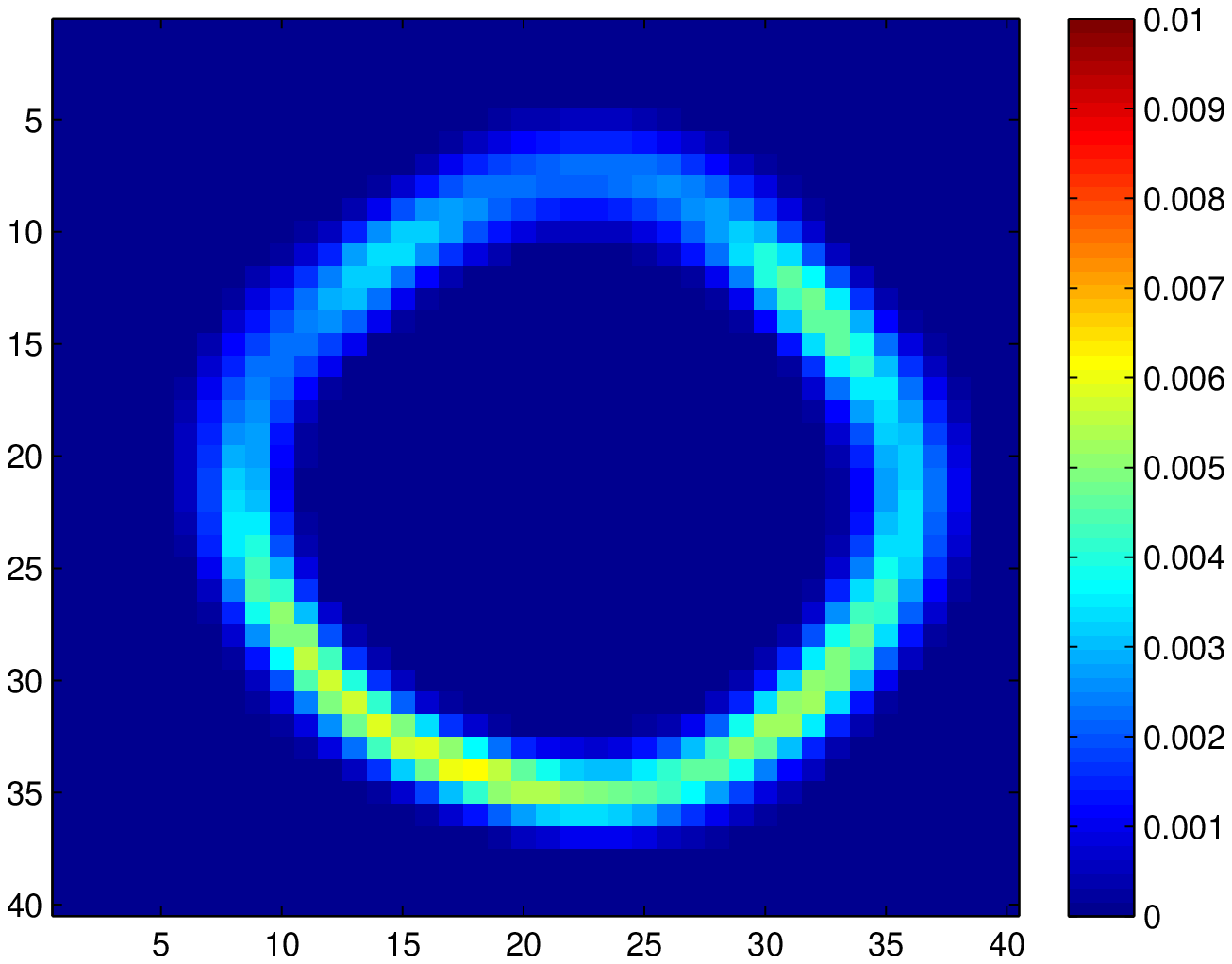}}
\subfigure[EASAL-1]{\label{fig:c}\includegraphics[scale=0.25]{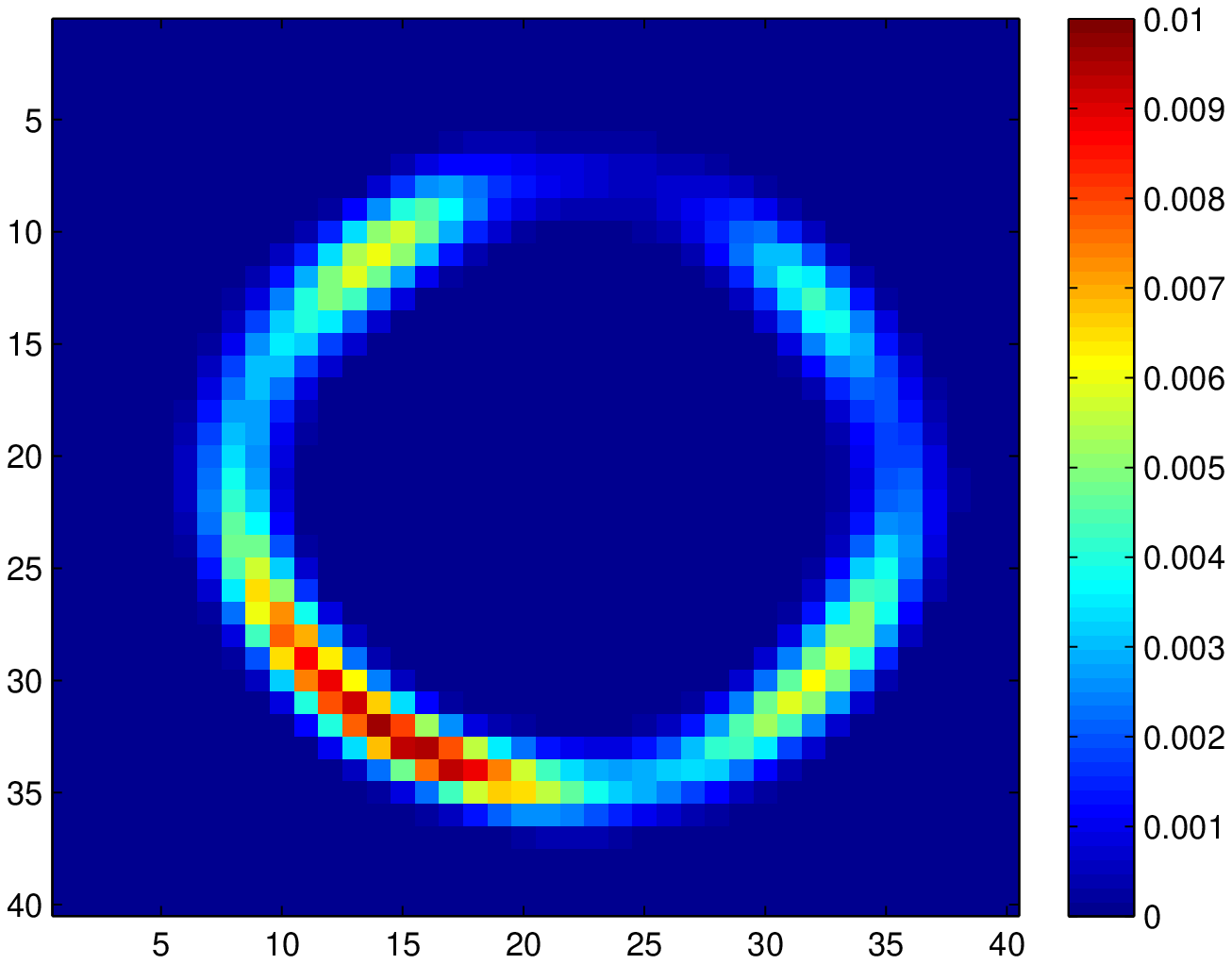}}
\subfigure[EASAL-2]{\label{fig:d}\includegraphics[scale=0.25]{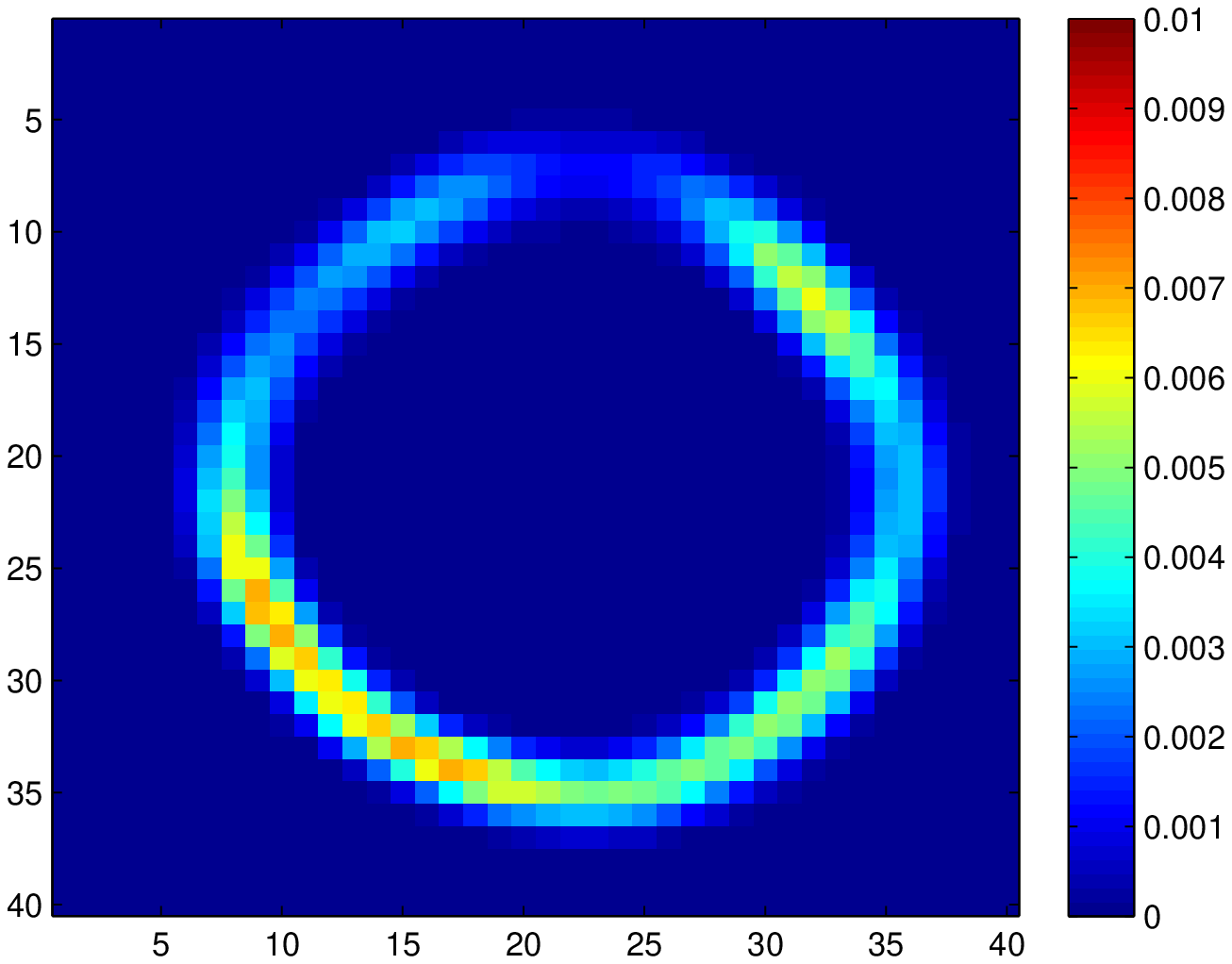}}
\subfigure[EASAL-3]{\label{fig:e}\includegraphics[scale=0.25]{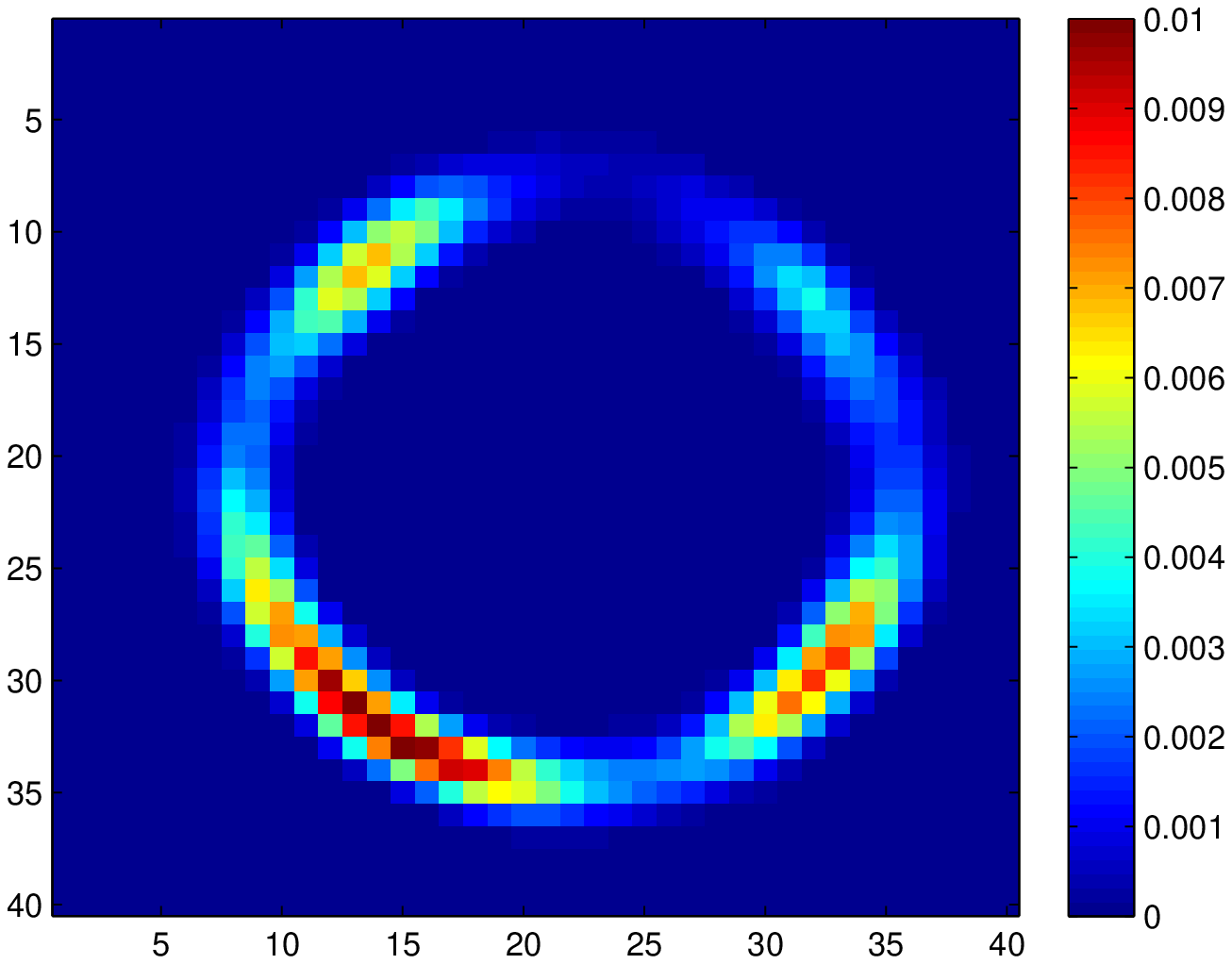}}
\subfigure[EASAL-Jacobian]{\label{fig:f}\includegraphics[scale=0.25]{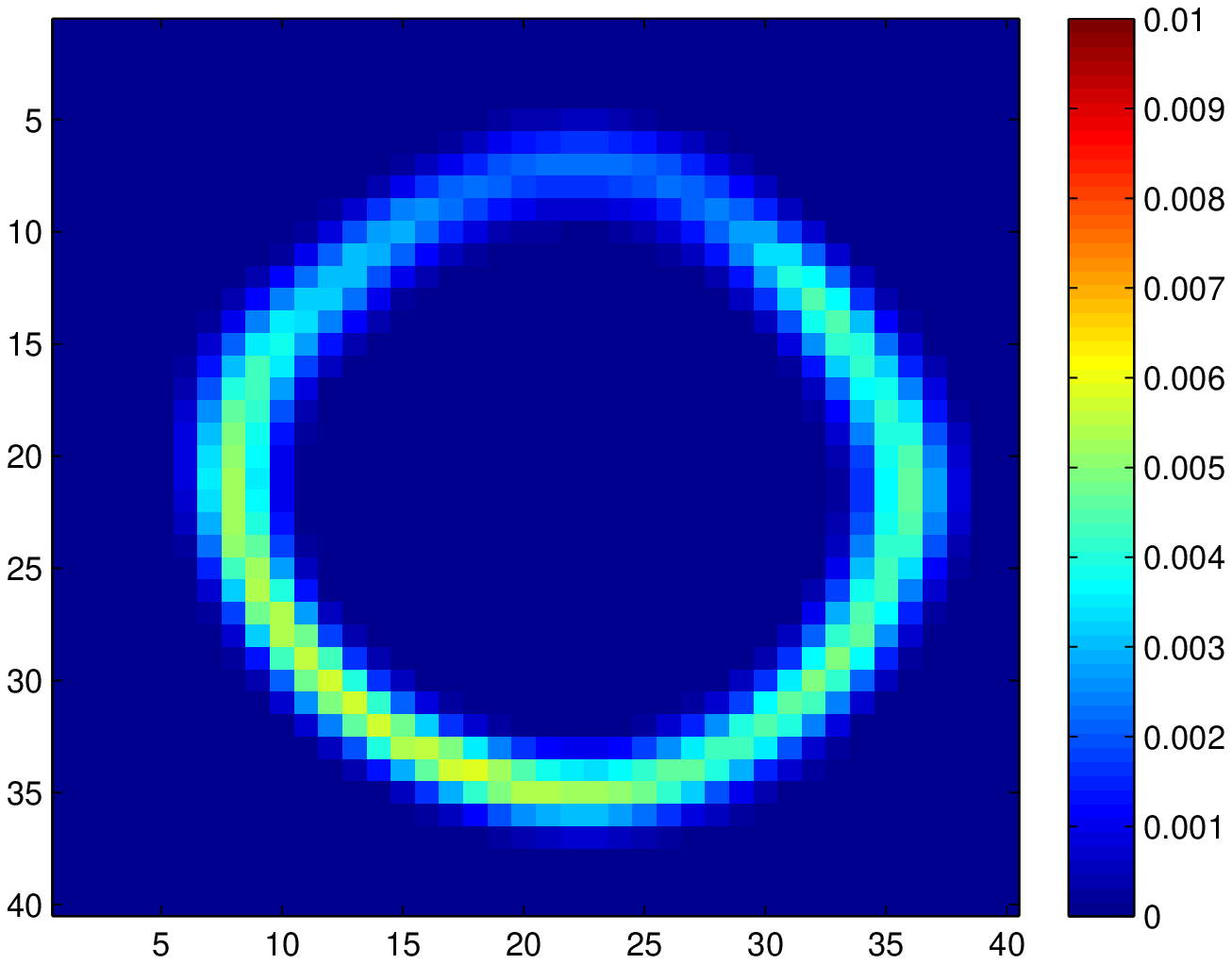}}
\caption{Projection in $\mathbb{R}^2$ of a configuration space for the example point-set shown
	in \figref{fig:3Input} as sampled by various methods. The
	projection is on the $xy$ coordinates of the centroid of the second
	point-set with the centroid of the first point-set fixed at the origin.
	The color scale on the right of each figure corresponds to the number
	of sampled points in a $\varepsilon$-sized cube centered around the grid
	point $(x, y)$. $\varepsilon$ is computed as described in the text \protect\cite{Ozkan2014MC}.}
\label{fig:projectionview}
\end{figure}

The experiments were run on an Intel i5-2540 machine and the variants of
EASAL were run on a Intel Core 2 Quad Q9450 @ 2.66 GHz. and a memory of 3.9 GB.
With this setup, EASAL-1 took 3 hours 8 minutes. EASAL-2 took 4 hours 24
minutes, EASAL-3 took 10 hours 20 minutes, and EASAL-Jacobian took 14 hours 22
minutes. The methods were compared based on a their sampling coverage of a
grid. The grid was set up to be uniform in the Cartesian configuration space
and its bounds along the $X$ and $Y$ axes were -20 to 20 Angstroms, and along
the $Z$ axes were -3.5 to 3.5 Angstroms.

The input in the experiment was as follows:

\begin{enumerate}
\item[(i)] The two point-sets in the form of two rigid helices. Note that this
is the special case of Problem (\cone, \ctwo) where the points are sphere centers.

\item[(ii)] The lower bound of the pairwise distance constraint, for
all sphere pairs $i, j$ belonging to different point-sets, $dist_{ij} > 0.8 \times (\rho_i
+\rho_j)$ where $i$ and $j$ are residues, $dist_{ij}$ is the distance between
residues $i$ and $j$, $\rho_i$ and $\rho_j$ are the radii of residue spheres $i$ and
$j$ respectively.

\item[(iii)] An optional global constraint 
is the inter helical angle between the principal axes of
the two input helices, $\theta < 30^{\circ}$. Here, $\theta = a~cos(uv)$
where $u$ and $v$ are the principal axis of each point-set, i.e., $u$ and $v$
are the dominant directions in which the mass is distributed, alternatively the
eigenvectors of the inertia matrix. 

\end{enumerate}

Over 43.5 million grid configurations were generated to ensure at least one pair
was an active constraint, i.e., $dist_{ij} < \rho_1 + \rho_2 + 0.9$. Out of these, around
86\% were discarded, leaving us with about 5.8 million `good' samples.

The methods were compared based on the following parameters.
\begin{itemize}
\item [-] The epsilon coverage: a measure of how many sample points are within
an $ \varepsilon$-sphere of each grid point. Since the ambient space has
dimension 6, $\varepsilon$ is set to
$(\text{number of grid points} / \text{number of sampling points})^{1/6} /2$.

\item [-] The coverage percentage, which is the percentage of the grid
$\varepsilon$-covered by the sampling algorithm. 

\item [-] The number of samples required to achieve the given
$\varepsilon$-coverage.

\item [-] The ratio percentage: Let $s_1$ be the number of samples in a specific but randomly chosen
3 dimensional region and $s_2$ be the number of samples in all 
ancestor regions with 1 active constraint that lead to the 3 dimensional region.  The
ratio percentage is $\frac{s_1}{s_2}\times 100$.

\end{itemize}

The best method should have the highest epsilon coverage and coverage
percentage with the fewest samples. As can be seen from
Table~\ref{table:coverage}, MC gives the best coverage but requires 100
million samples. By contrast EASAL-Jacobian gives about the same relative
coverage with one million samples (1\% of MC). EASAL-2 gives a very good
coverage of 92.42\% with only 40k samples (0.04\% of MC). EASAL-2 also has the
best ratio percentage beating even MC by a large margin.
\figref{fig:projectionview} shows a 2D projection of a configuration space as
sampled by various methods for the example point-set shown in
\figref{fig:3Input}. The projection is on the $xy$ coordinates of the
centroid of the second point-set with the centroid of the first point-set fixed
at the origin. Multigrid shows grid sampling where lower dimensional regions
are repeat sampled, which is desirable. More precisely, each grid point in a
$d$ dimensional region of the atlas with $6-d$ active constraints is weighted
by $6-d$. Notice that EASAL-Jacobian and EASAL-2 approximate Multigrid (target)
better than MC.

\subsection{Viral Capsid Interaction}

\begin{figure}
\includegraphics[scale = 0.3]{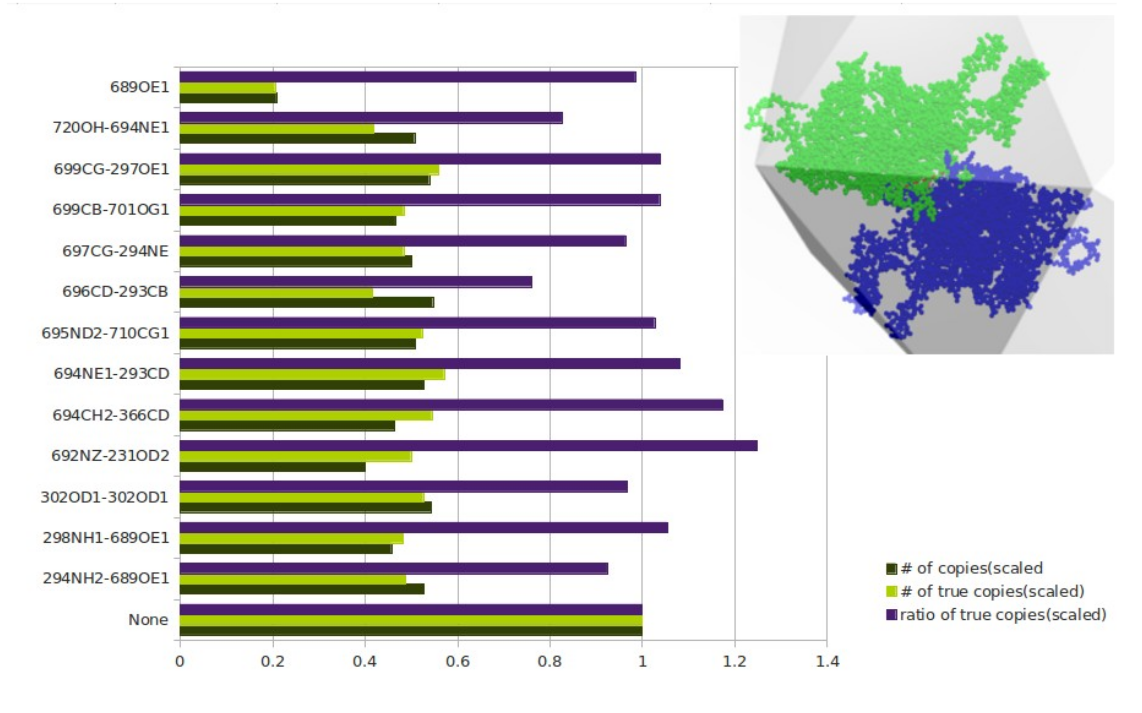}

\caption{Assembly of a dimer, 2-fold interface of
the icosahedral AAV2 virus capsid. Each row corresponds to removing a
particular residue pair. These are normalized to the bottom row where no
interaction is removed and shows respectively the total number of
zero-dimensional (rigid) configurations, the number of configurations close to
the successful interface assembly configuration, and their ratio.
\protect\cite{Wu2014Virus}.}
\label{fig:virus_comparison_dimer}
\end{figure}

\begin{figure}
\includegraphics[scale = 0.3]{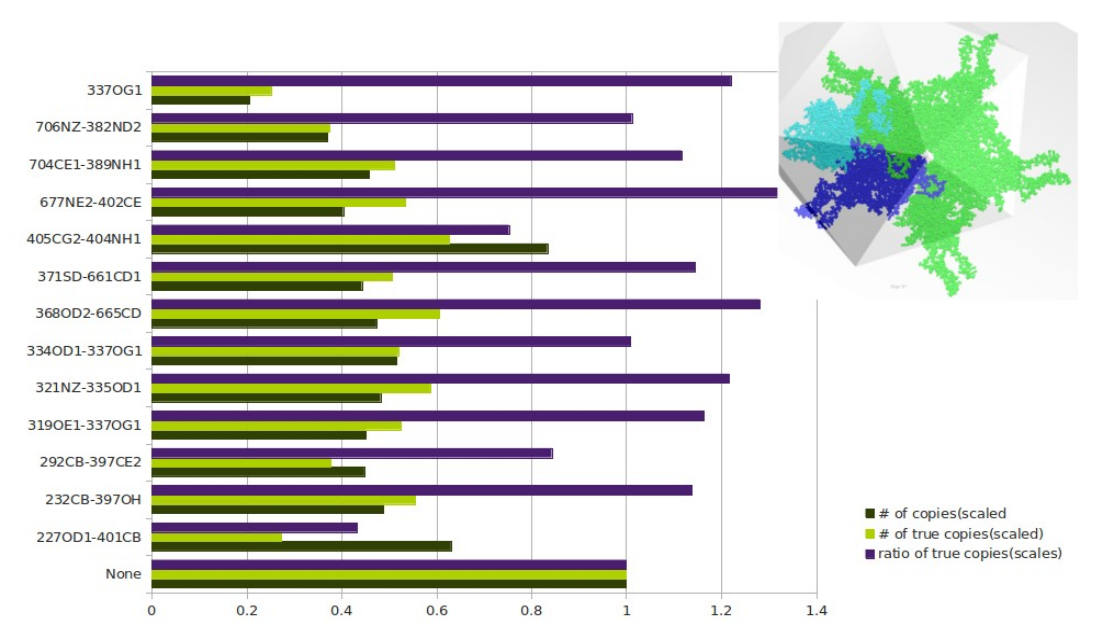}
\caption{Assembly of a pentamer, 5-fold interface
of the icosahedral AAV2 virus capsid. Each row corresponds to removing a
particular residue pair. These are normalized to the bottom row when no
interaction is removed and shows respectively the total number of
zero-dimensional or rigid configurations, the number of configurations close to
the successful interface assembly configuration, and their ratio.
\protect\cite{Wu2014Virus}.}
\label{fig:virus_comparison_pentamer}
\end{figure}

In this section we sketch results from \cite{Wu2014Virus}. EASAL has been
applied to study the configuration space of autonomous assembly into empty
shells of icosahedral T=1 viruses from nearly identical protein monomers
containing $n \ge 5000$ atoms. The robustness of such an assembly depends on
the sensitivity of free energy landscapes of inter-monomer interfaces to
changes in the governing inter-atomic interactions. The sensitivity towards
assembly disruption is generally measured by wet lab mutagenesis that disables
the chosen inter-monomer atomic interactions. \cite{Wu2014Virus} predicted this
sensitivity for the first time using EASAL to atlas the inter-monomer interface
configuration space, exploiting symmetries, and utilizing the recursive
decomposition of the large viral capsid assembly into an assembly pathway of
smaller assembly intermediates. The predictions were compared with the results
from the mutagenesis. Specifically, EASAL was used to predict the sensitivity
of 3 viral systems: Minute Virus of Mice (MVM), Adeno-Associated Virus (AAV2),
and Bromo-Mosaic Virus (BMV). For the case of AAV2, \figref
{fig:virus_comparison_dimer} and \figref {fig:virus_comparison_pentamer} show
the effect of removing a particular residue pair (the BMV results are not shown
here, \cite{unpublished}). Each row shows the total number of zero-dimensional
or rigid configurations, the number of configurations close to the successful
interface assembly configuration, and their ratio. Table \ref{table:virus}
shows comparison of the cruciality or sensitivity ranking thereby obtained to
the mutagenesis result. The highest ranked interactions output by EASAL were
validated by mutagenesis resulting in assembly disruption. The sensitivity
ranking of the dimer interface shows that all the residues marked crucial by
EASAL were confirmed as crucial by wet lab mutagenesis. The entries not listed
in the table, corresponding to non-crucial interactions, were either confirmed
as not crucial or there were no experiments performed for them. The sensitivity
ranking for the pentamer shows similar results, however experiments for some of
the residues marked by a question mark were not performed.

\begin{table}[h]

\begin{minipage}{0.45\linewidth}
\begin{tabular}{ccc}\hline
Residue1 & Residue2 & Confirmed\\\hline\hline
P293 & W694, P696 & Yes$^{*, \dag}$\\
R294 & E689, E697 & Yes $^{*, \dag, **}$\\
E689 & R298 & Yes $^{*, \dag}$\\
W694 & P293, Y397 & Yes $^{*, \dag}$\\
P696 & P293 & Yes $^{*, \dag}$\\
Y720 & W694 & Yes $^{*, \dag}$\\\hline
\end{tabular}
\subcaption{Sensitivity ranking: Dimer Interface\\}
\end{minipage}
\begin{minipage}{0.45\linewidth}
\begin{tabular}{ccc}\hline
Residue1 & Residue2 & Confirmed\\\hline\hline
N227 & Q401 & Yes $^{**}$\\
R389 & Y704 & ?\\
K706 & N382 & ?\\
M402 & Q677 & Yes $^{*, \dag}$\\
K706 & N382 & ?\\
N334 & T337,Q319 & ?\\
S292 & F397 & Yes $^{**}$\\\hline
\end{tabular}
\subcaption{Sensitivity ranking: Pentamer Interface\\}
\end{minipage}
\caption{Sensitivity ranking for the dimer and pentamer interface of AAV2. For
some residue pairs, marked by `?', there were no experiments performed and their 
cruciality is unconfirmed. $^*$ - \protect\cite{Rayaprolu15122013}, $^\dag$ - 
\protect\cite{mutagenesis}, $^{**}$ - \protect\cite{Wu:00} }
\label{table:virus}
\end{table}

%% file: sections/Architecture.tex
\section{Software architecture}
\label{architecture}

\begin{figure}
\centering
\includegraphics[width=\textwidth] {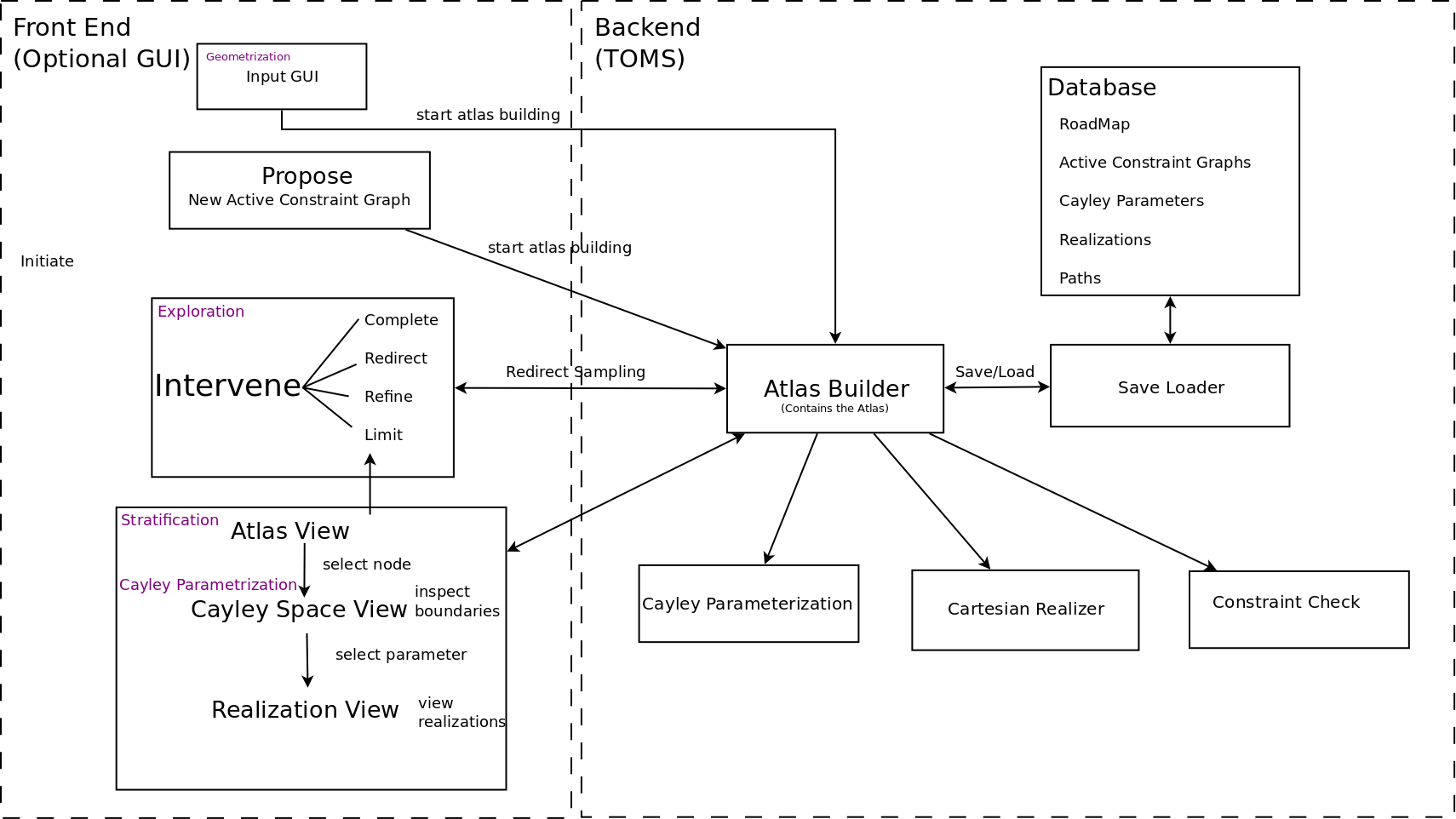}
\caption{Architecture of \EASAL.}
\label{fig:Architecture}
\end{figure}

The EASAL software has two versions. The TOMS submission contains only the
backend of EASAL, without GUI and with text input and output.  An optional GUI
(not part of TOMS submission) which can be used for intuitive visual
verification of the results, can be found at the EASAL bitbucket repository
\cite{easalSoftware}. \figref{fig:Architecture}, which shows the overall
architecture of EASAL, clearly demarcates these two versions.

The user initiates the sampling either by running just the backend in a
terminal or through the optional GUI (not part of TOMS submission). The
AtlasBuilder starts the sampling process by making a recursive call to the
`sampleAtlasNode' algorithm with the root node as the parameter. The Atlas
builder interacts with various components such as `Cayley parameterization',
`Cartesian Realization' and `Constraint Check' to help in the sampling process.
It uses the `SaveLoader' to save the generated atlas to the database.  All the
sampling information such as the atlas, active constraint graphs, Cayley
parameters and realizations are written to a database to avoid re-sampling.

When EASAL is initiated using the backend, the output is in text format.
The following are the output:
\begin{itemize}
\item The \emph{Roadmap}, which stores the atlas, i.e., a topologically
stratified set of sample feasible realizations or configurations of the two
rigid point sets. This can be found in the `RoadMap.txt' file in the data
folder.

\item The \emph{Node} files which contain sampling information, Cayley
parameter values, and realizations of the point sets. Each `Node*.txt' file
contains samples for a particular active constraint region.

\item The \emph{paths} file which contains the one degree of freedom motion
path between all pairs of lowest energy configuration regions. This can be
found in the `paths.txt' file in the data folder.

\item The \emph{path matrix}, which contains a path matrix where the rows and
columns correspond to 0D and 1D nodes. The $\{ij\}^{th}$ entry indicates the
number of paths between nodes $i$ and $j$. This can be found in the
`path\_matrix.txt' file in the data folder.
\end{itemize}

The optional GUI (not part of TOMS submission) can be used to visualize the
output of the backend. See Section 3.3.5 of the `Complete User Guide' located
in the bitbucket repository \cite{easalSoftware} for instructions.  The
optional GUI has three views: the \emph{atlas view}, the \emph{Cayley space
view} and the \emph{realization view}.  The atlas view shows the stratification
of the configuration space in the form of an atlas. In the atlas view, the user
can explore the atlas by intervening in the sampling process to either
complete, redirect, refine or limit the sampling.  The user can also propose
new constraints for active constraint graphs.  The Cayley space view shows the
user the Cayley configuration space of a node in the atlas.  In the Cayley
space view the user can view all the Cayley parameters and boundaries. In the
realization view, the user can view all the Cartesian realizations of the
selected node.  This view contains the \emph{sweep} feature which keeps one of
the point-sets fixed and draws the other point-set many times to trace out the
set of all feasible realizations.

%% file: sections/Conclusion.tex
\section{Conclusion}
\label{conclusion}
The EASAL software generates, describes, and explores key aspects of the
topology and geometry of the configuration space of point-sets in
$\mathbb{R}^3$. To achieve this, it uses three strategies, (i) EASAL partitions
the realization space into active constraint regions each defined by the set of
active constraints. The graph of active constraints called the active
constraint graph is then used for analysis using generic combinatorial rigidity
theory.  (ii) EASAL organizes the active constraint regions in a partial order
called an atlas which establishes a parent child relationship between active
constraint regions that generically differ by exactly one active constraint.
To build the atlas, EASAL starts from the interior of an active constraint
region and recursively finds boundaries of one dimension less.  (iii) To locate
the boundary region satisfying exactly one additional constraint, EASAL uses
the theory of Cayley convexifiability to map (many to one) a $d$-dimensional
active constraint region to a convex region in $\mathbb{R}^d$ called the Cayley
configuration space of the region.  This allows for efficient sampling and
search. In addition, it is efficient to compute the inverse map from each point
in the Cayley configuration space to its finitely many Cartesian realizations.
With EASAL we obtain formal guarantees for quantitative accuracy and running
times.  The EASAL software optionally provides a GUI which can be used for
intuitive visual verification of results.

In the context of molecular assembly, EASAL distinguishes assembly from other
processes such as folding in that assembly admits to Cayley convexification of
active constraint regions. More general methods like MC and MD, though
applicable to a wider variety of molecular modeling problems, do not make this
distinction and hence are not as efficient as EASAL in the context of molecular
assembly.  For the problem of assembly, EASAL (i) directly atlases and
navigates the complex topology of small assembly configuration spaces, crucial
for understanding free-energy landscapes and assembly kinetics; (ii) avoids
multiple sampling of configurational (boundary) regions, and minimizes rejected
samples, both crucial for efficient and accurate computation of configurational
volume and entropy and (iii) comes with rigorously provable efficiency,
accuracy and tradeoff guarantees. To the
best of our knowledge, no other current software provides such functionality.

The paper reviews the key theoretical underpinnings, major algorithms and their
implementation; outlines the main applications such as computation of
free energy and kinetics of assembly of supramolecular structures or of
clusters in colloidal and soft materials; and surveys select
experimental results and comparisons.